\documentclass[11pt,a4paper]{article}
\usepackage{epsfig}

\newcommand{\epm} {\ensuremath{{\rm e^{+} e^{-}}}}
\newcommand{\mupair} {\ensuremath{\mu^{+} \mu^{-}}}

\newcommand{\eett} {\ensuremath{\rm  e^+ e^- \rightarrow \tau^+ \tau^- }}

\newcommand{\tpm}{\ensuremath{\tau^{+} \tau^{-}}}
\newcommand{\taupair}{\ensuremath{\tau}-pair}

\newcommand{\pz}{\ensuremath{\pi^0}}
\newcommand{\pzs}{\ensuremath{\pi^0}'s}

\newcommand{\qqb}{\ensuremath{{\rm q} \overline{\rm q}}}

\newcommand{\decxA}{\ensuremath{\tau^- \rightarrow h^- \, \nu_\tau}}

\newcommand{\decxAP}{\ensuremath{\tau^- \rightarrow \pi^- \, \nu_\tau}}
\newcommand{\decxAK}{\ensuremath{\tau^- \rightarrow {\rm K^-} \, \nu_\tau}}

\newcommand{\decxB}{\ensuremath{\tau^- \rightarrow h^- \, \pi^0 \nu_\tau }}
\newcommand{\decxBP}{\ensuremath{\tau^- \rightarrow \pi^- \, \pi^0 \nu_\tau }}
\newcommand{\decxBK}
{\ensuremath{\tau^- \rightarrow {\rm K^-} \, \pi^0 \nu_\tau }}

\newcommand{\decxCC}
{\ensuremath{\tau^- \rightarrow h^- \, \geq 2 \pi^0 \nu_\tau }}

\newcommand{\decxD}{\ensuremath{\tau^- \rightarrow h^- \, 3 \pi^0 \nu_\tau }}
\newcommand{\decxE}{\ensuremath{\tau^- \rightarrow h^- \, 4 \pi^0 \nu_\tau }}

\newcommand{\decxKZ}
{\ensuremath{\tau^- \rightarrow h^- \, {\rm \overline{K}^0 \, X \,} \nu_\tau }}

\newcommand{\tauE}
{\ensuremath{\rm \tau^- \rightarrow e^- \, \overline{\nu}_e \, \nu_\tau }}
\newcommand{\tauM}
{\ensuremath{\rm \tau^- \rightarrow \mu^- \, \overline{\nu}_\mu \, \nu_\tau }}

\newcommand{\mtau}{\ensuremath{m_{\tau}}}

\newcommand{\beq}{\begin{equation}}
\newcommand{\eeq}{\end{equation}}
\newcommand{\bdm}{\begin{displaymath}}
\newcommand{\edm}{\end{displaymath}}
\newcommand{\bitm}{\begin{itemize}}
\newcommand{\eitm}{\end{itemize}}
\newcommand{\bfi}{\begin{figure}}
\newcommand{\efi}{\end{figure}}
\newcommand{\bce}{\begin{center}}
\newcommand{\ece}{\end{center}}

\newcommand{\MC}{Monte Carlo}

\newcommand{\NumOneProngs}{\ensuremath{158971}}
\newcommand{\NumThreeTracks}{\ensuremath{1903}}
\newcommand{\NumTwoTracks}{\ensuremath{5960}}

\newcommand{\reqmnts}{requirements}

\def\dedx{{\rm d}E/{\rm d}x}

\newcommand{\DataTauPairs}{\ensuremath{95364}}

\newcommand{\NeutClustLowUnscCut}{\ensuremath{0.5}}

\newcommand{\PizeroEoverPCut}{\ensuremath{1.0}}
\newcommand{\PzHiEnerUnscCut}{\ensuremath{9.0}}
\newcommand{\PzLowEnerUnscCut}{\ensuremath{2.2}}

\newcommand{\PzChiSqCut}{\ensuremath{9.0}}

\newcommand{\numzero}{\ensuremath{ 18547}}
\newcommand{\numone}{\ensuremath{ 40537}}
\newcommand{\numtwo}{\ensuremath{  6802}}
\newcommand{\totTau}{\ensuremath{190728}}
\newcommand{\reszerox}{\ensuremath{(11.98 \pm 0.13 \pm 0.16)\, \%}}
\newcommand{\resonex}{\ensuremath{(25.89 \pm 0.17 \pm 0.29)\, \%}}
\newcommand{\restwox}{\ensuremath{( 9.91 \pm 0.31 \pm 0.27)\, \%}}

\newcommand{\CorrelTwo}{\ensuremath{0.167}}
\newcommand{\CorrelOne}{\ensuremath{-0.430}}
\newcommand{\CorrelThree}{\ensuremath{-0.470}}

\newcommand{\UnivResult}{\ensuremath{1.018}}
\newcommand{\UnivResultErr}{\ensuremath{0.010}}
\newcommand{\NewUnivRes}{\ensuremath{1.003}}
\newcommand{\NewUnivResErr}{\ensuremath{0.005}}
\newcommand {\thn} {\mbox{$\tau^- \rightarrow h^- \nu_{\tau}$}}
\newcommand {\tpn} {\mbox{$\tau^- \rightarrow \pi^- \nu_{\tau}$}}
\newcommand {\tKn} {\mbox{$\tau^- \rightarrow \mathrm{K}^- \nu_{\tau}$}}
\newcommand {\hmn} {\ensuremath{{ h^-} \rightarrow \mu^- \bar{\nu}_{\mu}}}
\newcommand {\Kmn} {\ensuremath{{\rm K^-} \rightarrow \mu^- \bar{\nu}_{\mu}}}
\newcommand {\lpRPV}[1] {\mbox{$\lambda^{\prime}_{#1}$}}
\newcommand {\sdkR}     {\mbox{$\tilde{\mathrm{d}}^k_R$}}
\newcommand {\msdkR}    {\mbox{$m(\tilde{\mathrm{d}}^k_R)$}}
\newcommand {\msdkRsq}  {\mbox{$m^2(\tilde{\mathrm{d}}^k_R)$}}
\newcommand {\mmu}      {\mbox{$m_{\mu}$}}
\newcommand {\LsssWAthree}{0.09}
\newcommand {\LsssWAtwo}  {0.07}
\newcommand {\LsssOPthree}{0.15}
\newcommand {\LsssOPtwo}  {0.09}
\newcommand {\LsWAthree}{0.07}
\newcommand {\LsWAtwo}  {0.05}
\newcommand {\LsOPthree}{0.13}
\newcommand {\LsOPtwo}  {0.06}

\begin{document}
%
% ......... Title Page        ....................
\begin{titlepage}
\begin{center}{\large   EUROPEAN LABORATORY FOR PARTICLE PHYSICS
}\end{center}\bigskip
\begin{flushright}
       CERN-PPE/97-152   \\ 4 December 1997
\end{flushright}
\bigskip\bigskip\bigskip\bigskip\bigskip
\begin{center}{\huge\bf 
       Measurement of the one-prong hadronic tau branching ratios at LEP
}\end{center}\bigskip\bigskip
\begin{center}{\LARGE The OPAL Collaboration
}\end{center}\bigskip\bigskip
\bigskip\begin{center}{\large  Abstract}\end{center}
\noindent The branching ratios of the \decxA, \decxB\ and \decxCC\ 
decays have been measured 
using the 1991--1995 data recorded with the OPAL detector at LEP.
These branching ratios are measured simultaneously using three selection
criteria and are found to be 
\begin{center}
\begin{tabular}{l c c} \\
BR(\decxA)  & = & \reszerox \\
BR(\decxB)  & = & \resonex  \\
BR(\decxCC) & = & \restwox
\end{tabular}
\end{center}
where the first error is statistical and the second is systematic.
\bigskip\bigskip\bigskip\bigskip
\bigskip\bigskip
\begin{center}{\large
(To be submitted to Zeitschrift f\"{u}r Physik C) 
}\end{center}
\end{titlepage}
% ......... end of title Page ....................
%
% ......... author list ..........................
%

\begin{center}{\Large        The OPAL Collaboration
}\end{center}\bigskip
\begin{center}{
%begin authorlist
K.\thinspace Ackerstaff$^{  8}$,
G.\thinspace Alexander$^{ 23}$,
J.\thinspace Allison$^{ 16}$,
N.\thinspace Altekamp$^{  5}$,
K.J.\thinspace Anderson$^{  9}$,
S.\thinspace Anderson$^{ 12}$,
S.\thinspace Arcelli$^{  2}$,
S.\thinspace Asai$^{ 24}$,
S.F.\thinspace Ashby$^{  1}$,
D.\thinspace Axen$^{ 29}$
% ...
%end authorlist
}\end{center}\bigskip
\bigskip
%begin institutes
$^{  1}$School of Physics and Astronomy,
University of Birmingham,
Birmingham B15 2TT, UK
\newline
$^{  2}$Dipartimento di Fisica dell' Universit\`a di Bologna and INFN,
I-40126 Bologna, Italy
\newline
% ...
$^{  5}$Cavendish Laboratory, Cambridge CB3 0HE, UK
\newline
% ...
$^{  8}$CERN, European Organisation for Particle Physics,
CH-1211 Geneva 23, Switzerland
\newline
$^{  9}$Enrico Fermi Institute and Department of Physics,
University of Chicago, Chicago IL 60637, USA
\newline
% ...
$^{ 12}$Indiana University, Department of Physics,
Swain Hall West 117, Bloomington IN 47405, USA
\newline
% ...
$^{ 16}$Department of Physics, Schuster Laboratory, The University,
Manchester M13 9PL, UK
\newline
% ...
$^{ 23}$Department of Physics and Astronomy, Tel Aviv University,
Tel Aviv 69978, Israel
\newline
$^{ 24}$International Centre for Elementary Particle Physics and
Department of Physics, University of Tokyo, Tokyo 113, and
Kobe University, Kobe 657, Japan
\newline
% ...
$^{ 29}$University of British Columbia, Department of Physics,
Vancouver BC V6T 1Z1, Canada
\newline
% ...
\newline
%end notes
%
% ......... end of author list ...................
%
% ......... Introduction      ....................
%
\section{Introduction}

Precise tests of the Standard Model can be made using $\tau$ 
leptons~\cite{Pich_Tau96}.
Measurements of
the $\tau$ allow the study of the structure of the weak currents
and the universality of the couplings of the charged leptons to the gauge bosons,
and can be used to search for evidence of new physics.
Further, the $\tau$ is the only lepton to decay into hadrons, allowing
the study of the strong interaction.
For many of these studies an accurate knowledge of the 
properties of the $\tau$ is essential.
In this paper we report on a measurement of the branching ratios
of the \decxA, \decxB\ and \decxCC\ decays
\footnote{Charge conjugation is implied throughout this paper. The symbol
$h^-$ is used to indicate either $\pi^-$ or $\rm K^-$.} 
using the 1991--1995 data recorded with the OPAL detector at LEP.

The measurement of the one-prong hadronic branching ratios 
is made by selecting a sample of tau decays with one
track (one-prong) and then counting the number of \pzs\ in each decay.
The one-prong decays were subdivided into samples 
with $0$, $1$ or $\geq 2$ \pzs\
from which the branching ratios for the three signal 
channels were determined.
In Table~\ref{sig_chan} we list the $\tau$ decays that are included in each
signal channel.
We use the PDG definitions for these signal channels~\cite{PDG96}.
Note that not all modes were included in the \MC\ simulation.
No attempt to separate charged pions and kaons was made in this measurement.
The limited granularity of the OPAL electromagnetic calorimeter made it 
impossible to resolve unambiguously  the decays with two \pzs\ from those with 
three or more \pzs\ with the \pz\ identification algorithm used in this work,
hence only a measurement of the $ \geq 2 \pz $ branching ratio is presented.
\begin{table}[h]
\begin{center}
\begin{tabular}{l l l l} \hline
 Selection & Decay mode & Weight & Comment \\ \hline
\decxA   & $ \tau^- \rightarrow \pi^- \, \nu_\tau   $   &  &           \\ 
         & $ \tau^- \rightarrow {\rm K^-} \, \nu_\tau  $   &  &        \\ \hline
\decxB   & $ \tau^- \rightarrow \pi^- \, \pi^0 \, \nu_\tau   $  &  &   \\
         & $ \tau^- \rightarrow {\rm K^-} \, \pi^0 \, \nu_\tau $  &  & \\ \hline
\decxCC  & $ \tau^- \rightarrow \pi^- \, 2 \pi^0 \, \nu_\tau $  &  &   \\
         & $ \tau^- \rightarrow {\rm K^-} \, 2 \pi^0 \, \nu_\tau $ &  & \\
         & $ \tau^- \rightarrow \pi^- \, 3 \pi^0 \, \nu_\tau $  &  &   \\
         & $ \tau^- \rightarrow {\rm K^-} \, 3 \pi^0 \, \nu_\tau $  &  & Not modelled  \\
         & $ \tau^- \rightarrow h^- \, 4 \pi^0 \, \nu_\tau $  &  & Not modelled  \\
         & $ \tau^- \rightarrow \pi^- \overline{\rm K}^0 \nu_\tau $   & 0.157$^\dagger$ & \\
         & $ \tau^- \rightarrow {\rm K^-} \overline{\rm K}^0 \nu_\tau $  & 0.157$^\dagger$ & Not modelled \\
         & $ \tau^- \rightarrow \pi^- \overline{\rm K}^0 \pz \nu_\tau $  & 0.157$^\dagger$ & \\
         & $ \tau^- \rightarrow {\rm K^-} {\rm K}^0 \pz \nu_\tau $  & 0.157$^\dagger$ & Not modelled \\
         & $ \tau^- \rightarrow \pi^- {\rm K}^0 \overline{\rm K}^0 \nu_\tau $  & 0.0246$^\dagger$ & \\
         & $ \tau^- \rightarrow \pi^- \pz \, \eta \, \nu_\tau $  & 0.319$^*$ & \\ \hline
{\footnotesize $\dagger $ Only the $\rm K^0_S \rightarrow 2 \pz $ decay included.} & & & \\
{\footnotesize $*$ Only the $\rm \eta \rightarrow 3 \pz $ decay included.} & & & \\
\end{tabular}
\caption{\label{sig_chan} The various decay modes for each selection.
The weights reflect the fraction of each decay that contributes to the 
signal.
The weight is equal to unity when no number is explicitly given.
A number of the decay modes were not simulated in the Monte Carlo and
these are indicated in the last column.}
\end{center}
\end{table}

%
% ......... OPAL detector     ....................
%
\section{OPAL detector}

A detailed description of the OPAL detector can be found in 
Ref.~\cite{opal_det}.
A description of the features relevant for this analysis follows.

A high precision silicon microvertex detector surrounds the beam pipe.
This covers an angular region of $| \cos{\theta} | \leq 0.8$ and provides
hit information in the $x-y$ (and $z$ after 1992) directions\footnote
{The OPAL coordinate system defines the $+z$ axis 
in the $e^-$ beam direction. The angle $\theta$ is measured from the
$+z$ axis and $\phi$ is measured about the z axis from the +x axis 
which points to the centre of the LEP ring.}~\cite{opal_sidet}.
Charged particles are tracked in a central detector enclosed
inside a solenoid that provides a uniform axial magnetic
field of $0.435$T.
The central detector consists of three drift chambers:
a high resolution vertex detector, a large volume drift jet chamber
and the $z$-chambers.
The jet chamber records the momentum and energy loss of charged particles
over $ 98 \% $ of the solid angle and the $z$-chambers are used to 
improve the track position measurement in the $z$ 
direction~\cite{opal_cjdet}.

Outside the solenoid coil are scintillation counters  which
measure the time-of-flight from the interaction region and aid 
in the rejection of cosmic events.
Next is the electromagnetic calorimeter (ECAL) that is divided into barrel
($|\cos{\theta}|<0.82$) and end-cap ($ 0.81 < | \cos{\theta} | < 0.98 $) 
sections.
The barrel section is composed of $9440$ lead-glass
blocks each subtending approximately $\rm 40 \times 40 \, mrad^2$ and 
with a depth of
$24.6$ radiation lengths.

Beyond the electromagnetic calorimeter the iron of the solenoid return
yoke is segmented into layers and instrumented with limited streamer 
tubes as the hadron calorimeter (HCAL).
In the region $|\cos{\theta}|<0.81$ this detector typically 
has a depth of $8$ interaction lengths.
Beyond the hadron calorimeter is the muon chamber system, composed of 
four layers of drift chambers in the barrel region.

%
% ......... Event selection   ....................
%
\section{Event selection}
\label{ev_sel}

The results presented in this paper are based on the data taken during
the 1991-95 runs with the OPAL detector at LEP.
Approximately $90 \%$ of the data were taken at a centre-of-mass energy 
equal to the mass of the $Z^0$ boson $ M_{Z}$,
with the remaining data taken within $3$~GeV of $M_{\rm Z}$.

The \MC\ samples used in this analysis consist of $300\,000$ \taupair\ events 
generated at $\sqrt{s} = M_{\rm Z}$ and two samples of 
$100\,000$ events each generated respectively at $2$~GeV above and below
$M_{\rm Z}$.
The \MC\ samples were
generated with KORALZ 4.0~\cite{koralz} and TAUOLA 2.0~\cite{tauola} and then 
processed through the GEANT~\cite{geant} 
OPAL detector simulation~\cite{gopal}.
For this analysis an admixture of events from the \MC\ samples
generated above and below $M_{\rm Z}$ are added 
to the events generated at $M_{\rm Z}$ to reflect the 
distribution of centre-of-mass energies in the 1991--95 data set.

It has been found in a recent analysis~\cite{opal_had_paper} that the model
of K\"{u}hn and Santamaria (KS)~\cite{ks_model}, which is used in TAUOLA 2.0, 
does not satisfactorily describe the dynamics of the tau decay 
through the $ a_1 $ resonance.
The model of Isgur {\it et al.} (IMR)~\cite{imr_model}, although also not providing a 
completely satisfactory description of the 
$ \tau^- \rightarrow a_1^- \nu_{\tau} $ decay, was
found in~\cite{opal_had_paper} to be in better agreement with the data than
the KS model.
This improvement was found to be due mainly to the inclusion in
the IMR model of a polynomial background term, which accounts for
$ (13.8 \pm 2.4) \% $ of the total decay.

For this analysis, we have therefore chosen to apply weights to
the \MC\ generated events so that they are 
distributed dynamically like the IMR model description.
The weights applied are based on the values of $Q^2$, $s_1$, and $s_2$ 
at the tree level, where $Q^2$ is the invariant mass squared of
the $ \pi^- \, 2\pi^0 $ system.  The Dalitz plot variables $s_1$ and $s_2$ are
defined in terms of the pion 4-momenta as $ s_1 = (p_2 + p_3)^2 $
and $ s_2 = (p_1 + p_3)^2 $, with the labels chosen such that $ p_3 $
refers to the charged pion.
The weighting has been accomplished by generating normalized three 
dimensional arrays of decay rate $ \Gamma(Q^2,s_1,s_2) $ for each 
of the two models, and taking the bin-by-bin quotient as the weight 
factor.
The array grid size was chosen to be $ 0.05 \, {\rm GeV^2} $ for 
each of $ Q^2 $, $ s_1 $, and $ s_2 $.  
The model parameters were taken from~\cite{opal_had_paper} for
the IMR model, while KS model parameters, taken from~\cite{ks_model}, 
were those used in TAUOLA 2.0.
A smoothing algorithm was applied to the
IMR array to force the polynomial background term to zero
smoothly at the $ s_1 $ and $ s_2 $ physical boundaries, but in such 
a way as to preserve the measured $ Q^2 $ dependent shape and 
fractional contribution of the polynomial background term~\cite{opal_had_paper}.

The event selection starts by identifying \eett\ events,
then $\tau$ decays with one track are identified and clusters are
formed in the electromagnetic calorimeter.
The \pzs\ in each one-prong jet are reconstructed and
background $\tau$ decays are rejected.

%
% ......... Tau Pair selection   ....................
%
\subsection{Tau pair selection}
\label{tau_sel}

The \taupair\ sample is created by selecting events in the barrel region 
of the detector
with two back-to-back cones or {jets}~\cite{jet_defin}, of 
half-angle $35^\circ$~degrees.
Each event is required to have
\bdm
E_{\rm ECAL} + E_{\rm track} > 0.01 \, E_{\rm CM} \, ,
\edm
\bdm
\overline{ | \cos{\theta} | } < 0.68 \, ,
\edm
where $E_{\rm track}$ is the scalar sum of the momenta of the tracks,
$E_{\rm ECAL}$ is the total energy of the clusters in the 
electromagnetic calorimeter,
$E_{\rm CM}$ is the centre-of-mass energy of the \epm\ beams and
$\overline{ | \cos{\theta} | }$ is the average value of 
${ | \cos{\theta} | }$ for the two jets.

Cosmic and beam related backgrounds are rejected by placing requirements on the 
time-of-flight detector.
Additional requirements are needed to separate the \tpm\ events from other
two fermion background ($\epm \rightarrow {\rm f} \overline{\rm f}$) events:

\begin{itemize}

\item
Multihadronic events ($\epm \rightarrow {\rm q} \overline{\rm q}$) at the LEP energies
are characterized by large track and cluster multiplicities.
These events are rejected by requiring
at least two and not more than six tracks  
and not more than ten electromagnetic calorimeter clusters.

\item
Bhabha events ($ \epm \rightarrow \epm $) are characterized by two 
back-to-back
high energy charged particles that deposit close to centre-of-mass energy 
 in the electromagnetic calorimeter.
Bhabha events are rejected by requiring the 
\taupair\ candidates to have $ E_{\rm ECAL} \leq 0.8 E_{\rm CM}$ or
$ E_{\rm ECAL} + 0.3 E_{\rm track} \leq E_{\rm CM}$.

\item
Muon pair events ($ \epm \rightarrow \mupair $) are identified as two
high momentum back-to-back tracks 
that leave little energy in the electromagnetic calorimeter.
These events are removed if the tracks
have associated activity in the muon detectors or hadronic calorimeter
and $ E_{\rm ECAL} + E_{\rm track} > 0.6 E_{\rm CM}$.

\end{itemize}

In addition to the background from two fermion events, there are
also two-photon events
($\epm\rightarrow(\epm)X$, where $X = \, $ \epm, \mupair, \tpm, \qqb)
that must be rejected.
Two photon events leave little energy in the detector as the 
$\rm e^+$ and $\rm e^-$ particles are emitted at angles close to the beam
and are often undetected.
In addition, the detected particles tend to have a large acollinearity
angle.
The acollinearity angle, $\theta_{\rm acol}$, is defined to be the complement of
the angle between the two tau jets in the event.
These events are rejected by requiring
\bdm
\rm
\theta_{\rm acol} \leq 15^\circ \, ,
\edm
\bdm
E_{\rm vis} \geq 0.03 E_{\rm CM} \, ,
\edm
where 
%$\rm \theta_{\rm acol} $ is the acolinearity between the two jets and
$E_{\rm vis}$ is the sum of the visible energies of the jets
(taken for each jet as the maximum of the sum of the track and electromagnetic 
calorimeter cluster energies).
If $ E_{\rm vis} \leq 0.20 E_{\rm CM} $ then events are rejected 
if they satisfy
\bdm
P^T_{\rm tracks} > 2.0 \, {\rm GeV}\, 
{\rm or } \,\, 
P^T_{\rm ECAL} > 2.0 \, {\rm GeV}  \, ,
\edm
where $ P^T_{\rm tracks}$ ($P^T_{\rm ECAL}$) is
the vector sum of momentum (energy) of all tracks 
(electromagnetic calorimeter clusters) in the transverse direction.

The tau selection applied to all data 
yields \DataTauPairs\ \taupair\ candidates.
The non-tau background contributions in the sample have been investigated
in reference~\cite{PPE95_142} and are shown in Table~\ref{non_tau_bkgrd}.
\begin{table}
\begin{center}
\begin{tabular}{l c} \hline
Background                           & Contamination (\%) \\ \hline
$ \epm \rightarrow \mupair $         & $ 0.72 \pm 0.05 $ \\
$ \epm \rightarrow \epm $            & $ 0.41 \pm 0.07 $ \\
$ \epm \rightarrow \qqb$             & $ 0.28 \pm 0.04 $ \\ 
$ \epm \rightarrow (\epm) \epm     $ & $ 0.07 \pm 0.02 $ \\
$ \epm \rightarrow (\epm) \mupair  $ & $ 0.08 \pm 0.02 $ \\ \hline \hline
Total                                & $ 1.56 \pm 0.10 $ \\ \hline 
\end{tabular}
\caption{\label{non_tau_bkgrd} Non-tau background in the
\taupair\ sample.}
\end{center}
\end{table}
The $ \epm \rightarrow \qqb$ background has been 
re-evaluated by comparing the
number of clusters in the electromagnetic calorimeter with
those predicted by the $\tau$ Monte Carlo.  
The $ \epm \rightarrow \qqb$ events that
enter the $\tau$ sample tend to have a larger number of clusters
than \taupair\ events.

%
% ......... One-prong selection   ....................
%
\subsection{One-prong selection}
\label{one_prong_sel}

Jets with one track are identified as one-prong decays.
Jets with 2 or 3 tracks may also be identified as one-prong decays
if the additional tracks are associated with a photon conversion.
The track not identified as a conversion electron is henceforth called 
the primary track.
The photon conversion algorithm used in this analysis is described
in \cite{idncon}.
A total of \NumOneProngs\ one prong jets are selected of which
\NumTwoTracks\ are jets with 2 tracks and \NumThreeTracks\ are
jets with 3 tracks.

%
% ......... Clustering Algorithm  ....................
%
\subsection{Clustering algorithm}
\label{pi0_selection}

A fine clustering algorithm~\cite{v_clust_alg} is used to identify the 
particles in the one-prong $\tau$ decays.
The fine clustering algorithm limits the cluster size in the
electromagnetic calorimeter
to be $2 \times 2$ blocks in $\theta$ and $\phi$.
Clusters adjacent to other clusters may have fewer blocks.
Both data and \MC\ show that, on average, $99 \%$ of the 
energy of an electron and $95 \%$ of the energy of a charged pion that is 
deposited in the lead-glass calorimeter is contained
in the $2 \times 2$ cluster.

The electromagnetic calorimeter clusters are matched to the 
tracks using a $\chi^2$
significance parameter
in $\theta$ and $\phi$ that is weighted by the uncertainty in 
the track position
and the uncertainty in the cluster centroid.
An electromagnetic calorimeter cluster that is not associated to a 
track and that has energy
($ E_{\rm NC}$)
greater than $\NeutClustLowUnscCut$~GeV is classified as a {neutral cluster}.
The energy of the \MC\ electromagnetic calorimeter clusters has been 
smeared so that the energy resolution 
in the data and \MC\ are approximately equal.
Figure~\ref{nnclus2} shows the distribution of the
number of neutral clusters per one-prong jet
\footnote{The \MC\ distributions shown in all figures are 
normalized to the number of events after the \taupair\ selection.
The $\tau$ branching ratios used are those published by the 
PDG~\cite{PDG96} except for
the signal channels where the results of this analysis are used.}.

In about 1\% of $\tau$ decays there is a neutral cluster 
in the jet that is created by a radiative photon.
If the invariant mass of the primary track and any cluster
is greater than $3.0$~GeV, then that cluster is not considered
a neutral cluster.
Figure~\ref{rad_ener} shows  the distribution of the energy of neutral 
clusters removed by this requirement.

%
% ......... Pizero Reconstruction  ....................
%
\subsection{\pz\ identification}
\label{pi0_ident}

The \pz\ decays $98.8 \%$ of the time into two photons with the remaining 
$1.2 \%$ of the decays into an $\epm \gamma$ final state (Dalitz decay).
The algorithm used to identify \pzs\ is applied in four sequential steps:
\begin{enumerate}

\item Any neutral cluster in the jet with $ E_{\rm NC} > \PzHiEnerUnscCut $~GeV 
is identified as a \pz.
In the selected samples of \decxB\ and \decxCC\ jets, approximately
$ 55 \%$ and $ 34 \%$ of the $\pi^0$'s, respectively, 
are identified by this criterion.

\item Pairs of neutral clusters, or a neutral cluster and photon conversion,
each with neutral cluster or photon conversion
energy less than \PzHiEnerUnscCut~GeV, are candidates to form a \pz.
The pair is considered a \pz\ candidate if its energy 
is at least $3$~GeV and its invariant mass 
($m_{\gamma \gamma}$) is consistent with the \pz\ 
mass using a $\chi^2_{\pz}$ requirement.
The $\chi^2_{\pz}$ variable is defined as 
\bdm
\chi^2_{\pz} \equiv \frac{(m_{\gamma \gamma} - m_{\pz})^2}
                         {\sigma_{m_{\gamma \gamma}}^2} \, ,
\edm
where $m_{\pz}$ is the mass of the \pz\ meson and 
$\sigma_{m_{\gamma \gamma}}$ is the calculated uncertainty on 
$m_{\gamma \gamma}$.
A pair is considered a \pz\ if $ \chi^2_{\pz} < \PzChiSqCut $.

The number of \pzs\ formed in each jet using this method is not limited.
If there is an ambiguity between neutral clusters and photon conversion,
the combination that gives the best $\chi^2_{\pz}$ is chosen.
Figure~\ref{mass_gg} shows the invariant  mass distribution of 
two neutral clusters before and after 
this selection criterion is applied.
In the selected samples of \decxB\ and \decxCC\ jets, approximately
$ 17 \%$ and $ 14 \%$ of the $\pi^0$'s, respectively, 
are identified by this criterion.

\item 
Any remaining neutral clusters with 
$ E_{\rm NC} > \PzLowEnerUnscCut $~GeV 
are classified as \pzs.
In the selected samples of \decxB\ and \decxCC\ jets, approximately
$ 27 \%$ and $ 33 \%$ of the $\pi^0$'s, respectively, 
are identified by this criterion.

\item 
Frequently the \pz\ cannot be resolved from a track.
If the cluster associated to the track satisfies both
\bdm
{ E/p > \PizeroEoverPCut} \, ,
\edm
\bdm
{\rm and} \; \; \;
{ E - 0.3p > \PzLowEnerUnscCut \, {\rm GeV} } \, ,
\edm
then we consider the cluster to be an overlap \pz.
The energy of the \pz\  is  estimated to be $ E - 0.3p$ 
where $E$ is the energy of the cluster (charged hadron plus \pz).
The energy deposited by the charged hadron in the electromagnetic 
calorimeter is on average one-third of its momentum ($0.3p$).
These \pzs\ are only permitted in jets where one or more \pzs\ are 
identified by any of cases (1) to (3) above. 
If one or more conversion tracks point to the electromagnetic calorimeter cluster associated
to the primary track then the cluster energy ($E$) is modified by subtracting
the momenta of the conversion track(s).
In the selected samples of \decxCC\ jets, approximately
$ 19 \%$ of the \pzs\ are identified by this criterion.
\end{enumerate}

Each one-prong jet is classified as either \decxA, \decxB\ or \decxCC.
Figure~\ref{num_pi0} shows the distributions of
the number of \pzs\ per jet and the types of \pzs\ 
identified by this algorithm after background rejection (described below).
The energy distribution of \pzs\ in \decxB\ and \decxCC\ jets is shown in 
figure~\ref{en_pi0_x}.

%
% ......... Background Rejection  .....................
%
\subsection{Background rejection}
\label{back_rej}

A number of additional requirements are applied to remove residual backgrounds.
In particular, the \decxA\ sample has contamination from \tauE\ and \tauM\
decays. This contamination is reduced by requiring
\bdm
E/p < 0.75 \, ,
\edm
\bdm
p/E_{\rm beam} > 0.05 \, ,
\edm
\bdm
N^{\rm layers}_{\rm MB} = 0 \, ,
\edm
where $ E_{\rm beam} $ is the energy of the LEP beam and
$ N^{\rm layers}_{\rm MB}$ is the number of layers hit in the 
muon barrel detector.
In addition, backgrounds from $\epm \rightarrow \epm$ and 
$\epm \rightarrow \mupair$ events in the \decxA\ sample
are rejected by removing events 
where the acoplanarity angle between the two jets in the event
is less than $0.003$~radians and
the primary tracks in each jet have $ p > 30$~GeV.
The acoplanarity angle is defined to be the complement of the angle
between the two jets in the transverse plane of the event.

In the \decxB\ and \decxCC\ samples the main backgrounds
are due to misidentification of other signal channels.
In the \decxB\ and the \decxCC\ samples, the background is reduced 
by requiring the invariant jet mass to be less than $2.0$~GeV.
In the \decxB\ sample, the invariant jet mass is also required
to be greater than $0.4$~GeV.
The invariant mass distributions of the \decxB\ and \decxCC\ samples 
are shown in figure~\ref{inv_masses_1}.

A number of \decxB\ decays are mistakenly selected into the \decxCC\ sample.
The total energy of the \pzs\ from the \decxB\ decays is relatively
low and this contamination is reduced by requiring
\bdm
\frac{E_{\pz}^{\rm total}}{p} > 0.8 \, ,
\edm
where $E_{\pz}^{\rm total}$ is the energy sum of all \pzs\ 
identified in the jet
and $p$ is the momentum of the primary track.

%
% ......... Background Estimation .....................
%
\section{Estimation of $\tau$ backgrounds}
\label{back_est}

The backgrounds from \tauE\ and \tauM\ decays 
in each sample are measured using distributions
previously unused in the selection process.
A region in each distribution that is dominated by the background
is used to make the measurement.
The data and \MC\ distributions in the background-dominated region are
compared and any deviations between the two are assumed to be caused
by the background in question.
If the ratio of data to \MC\ is different from unity then the 
\MC\ predicted $\tau$ background
is rescaled in the background calculation.
The error of this correction factor is the combined statistical
errors of the data and \MC.
The uncertainties on the correction factors are included in the background
errors (including the case where the correction factor is unity).

The \tauE\ background is estimated using the energy 
loss ($\dedx$) in the jet drift chamber.
Figure~\ref{dedx_0pi0} shows the $\dedx$ distribution in the data 
and \MC\ for jets with one \pz\ and $p < 5$~GeV.
The ratios of data to \MC\ \tauE\ background jets
are calculated using jets that have $ 9.25 < \dedx < 12.0 $~keV/cm.
The ratio obtained in each sample is used as a correction factor 
and is shown in Table~\ref{tau_back_table}.

\begin{table}
\begin{center}
% --------------------------------------------------------------
% \input{./inputs/background_table} Nov 24 Version
\begin{tabular}{l c c} \\ \hline
\multicolumn{1}{c}{Selection} & \multicolumn{2}{c}{Background}        \\ \hline
          & \tauE\            & \tauM\            \\ \hline
\decxA\   & $ 1.30 \pm 0.18 $ & $ 1.39 \pm 0.26 $ \\
\decxB\   & $ 1.04 \pm 0.06 $ & $ 0.91 \pm 0.28 $ \\
\decxCC\  & $ 0.80 \pm 0.28 $ &                   \\ \hline
\end{tabular}
% --------------------------------------------------------------
\caption{\label{tau_back_table} Correction factors used to scale 
the \tauE\ and \tauM\ backgrounds.}
\end{center}
\end{table}

The \tauM\ background in the \decxA\ sample is measured by creating 
a sample of \tauM\ jets in both data and \MC.
The efficiency of the muon chamber requirements to reject \tauM\ is 
measured using these samples and the ratio of the efficiencies is 
used as the \MC\ scaling factor for \tauM\ background in the \decxA\ sample.
The \tauM\ contamination in the \decxB\ sample is 
measured by identifying tracks with hits in the muon chambers, since
the muon chambers were not used in this selection.
The ratio of data to \MC\ is used as the scaling 
factor for \tauM\ background in the \decxB\ sample.

Table~\ref{corr_backgrd} shows the estimated backgrounds in each selection.
In addition to the \tauE\ and \tauM\ decays, there is 
background from \decxKZ\ decays where X is any number of neutral
particles in the final state.
There is also background from other tau decays,
labelled as ``Other background" in Table~\ref{corr_backgrd}, which
are mainly tau decays to $\eta$ ($ 27 \% $) and $\omega$ modes ($ 60 \%$), 
with lesser contributions 
coming from decays with three charged hadrons in the final state ($ 13 \% $).
Both the \decxKZ\ and other tau decays are measured from \MC\ information.
The errors are calculated from the uncertainties in the efficiency matrix and
the background branching ratio from the PDG~\cite{PDG96}.
The background uncertainties in each selection, shown
in Table~\ref{corr_backgrd}, are used in calculating the systematic errors.
An additional modelling uncertainty, described in Section~\ref{syst_err},
is added to the \decxKZ\ decays.

\begin{table}
\begin{center}
% --------------------------------------------------------------------
% \input{./inputs/Bkgrd_3} Nov 24 version
\begin{tabular}{l r r r} \\ \hline
Background & \multicolumn{3}{c}{Selections} \\ \hline
 & \decxA & \decxB & \decxCC \\ \hline
\tauE\ & $  1.43 \pm 0.20 \% $ & $  1.59 \pm 0.11 \% $ & $  1.67 \pm 0.59 \% $ \\
\tauM\ & $  1.57 \pm 0.29 \% $ & $ 0.68 \pm 0.21 \% $ & $ 0.02 \pm 0.04 \% $ \\
\decxKZ\ & $  1.80 \pm 0.14 \% $ & $  1.57 \pm 0.11 \% $ & $  1.81 \pm 0.15 \% $ \\
Other background  & $ 0.07 \pm 0.02 \% $ & $ 0.53 \pm 0.03 \% $ & $  3.32 \pm 0.24 \% $ \\ \hline \hline
Total & $  4.88 \pm 0.38 \% $ & $  4.37 \pm 0.26 \% $ & $  6.82 \pm 0.65 \% $ \\ \hline
\end{tabular}
% --------------------------------------------------------------------
\caption{\label{corr_backgrd} The corrected backgrounds as a percentage
of each selection.}
\end{center}
\end{table}
%
% ......... BR Calculation    ................................
%
\section{Branching ratio calculation}
\label{br_rat_cal}

The branching ratios are calculated using the information from each selection
simultaneously.
Each selection can be expressed in terms of the efficiency for 
detecting each decay mode, the branching ratio of each mode and the number
of events selected in the data.
For each selection i, the equation is written as
\beq
\label{br_1}
\epsilon_{i1} B_1 + \epsilon_{i2} B_2 + \epsilon_{i3} B_3
+ \sum_{k=4}^{M} \epsilon_{ik} B_k =
\frac{N_i}{N_{\tau}(1-f^{non-\tau})}
\eeq
where $\epsilon_{ij}$ ($j=1,3$) are the efficiencies for selecting signal 
$j$ using selection $i$ and
$\epsilon_{ik}$ ($k=4,...$) are the efficiencies for selecting
the $\tau$ background modes using selection $i$.
$B_j$ ($j=1,3$) are the branching ratios of the signal channels and
$B_k$ ($k=4,...$) are the branching ratios of the backgrounds.
$N_i$ is the number of data events that pass the selection $i$,
$f^{non-\tau}$ is the fraction of non-tau events in the \taupair\ sample
and $N_\tau$ is the total number of data $\tau$'s that pass the 
\taupair\ selection.
The selection efficiencies ($\epsilon_{ij}$) for both signal and background
are determined from \MC\ with the main backgrounds corrected
using data distributions (described in section~\ref{back_est}).
The $\tau$ background branching ratios are taken from the PDG~\cite{PDG96} 
and $f^{non-\tau}$ is shown in Table~\ref{non_tau_bkgrd}.

The solution of the three simultaneous equations gives the 
branching ratios in the tau-selected sample.
These branching ratios are then corrected to account for 
the biases introduced to
the \taupair\ sample by the \taupair\ selection.
These factors are $0.989 \pm 0.002$, $1.019 \pm 0.001$ and $1.013 \pm 0.002$
for the \decxA\, \decxB\ and \decxCC\ decays, respectively.

%
% ......... Results    ................................
%
\section{Results}
\label{results}

The \taupair\ selection identifies $\totTau$ $\tau$ candidates.
The selections for \decxA, \decxB\ and \decxCC\ yield 
$\numzero$, $\numone$ and $\numtwo$ events respectively.
The efficiencies for detecting these signals are listed in 
Table~\ref{signal_effic_1}.
\begin{table}
\begin{center}
% -------------------------------------------------------------------
% {\input{./inputs/EffMat2.tex}} Nov 24 version
 \begin{tabular}{l c c c} \hline
 Selection & \multicolumn{3}{c}{Selection efficiency from MC}
 \\ \hline
  & \decxA & \decxB & \decxCC \\ \hline
$ 0 \pi^0 $ & $0.5502 \pm 0.0024$ & $ 0.1003 \pm 0 .0010$ & $ 0.0249 \pm 0 .0008$ \\ 
$ 1 \pi^0 $ & $ 0.0455 \pm 0.0010$ & $ 0.5985 \pm 0 .0016$ & $ 0.4273 \pm 0 .0025$ \\ 
$ \geq 2 \pi^0 $ & $ 0.0005 \pm 0.0001$ & $ 0.0237 \pm 0 .0005$ & $ 0.2730 \pm 0 .0023$ \\ 
  \hline 
 \end{tabular}
% -------------------------------------------------------------------
\caption{\label{signal_effic_1}
Efficiencies for identifying the signals for each selection.
The errors on these efficiencies are based on \MC\ statistics only.}
\end{center}
\end{table}
The efficiency for detecting \decxA\ is an average of the efficiencies for 
selecting \decxAP\ and \decxAK\ weighted by their relative branching
ratios.
Similarly, the efficiency for detecting \decxB\ is an average of the 
efficiencies for selecting \decxBP\ and \decxBK\ weighted by their 
relative branching ratios.
The same method is used for the \decxCC\ efficiency where the decay modes
are listed in Table~\ref{sig_chan}.

The backgrounds in
the \decxA, \decxB\ and \decxCC\ samples are given in Table~\ref{corr_backgrd}.
The branching ratios for \decxA, \decxB, and \decxCC\ are calculated
using the method described in section~\ref{br_rat_cal} and
the results are shown in Table~\ref{br_ratio_results}.
\begin{table}
\begin{center}
% -------------------------------------------------------------------
% {\input{./inputs/BrRatio3.tex}} Nov 24 version
 \begin{tabular}{ l r } \hline 
 Channel & \multicolumn{1}{c}{Branching Ratio ($\%$)} \\ \hline
\decxA  &  $ \rm {11.98 \pm 0.13 \pm 0.16} $ \\
\decxB  &  $ \rm {25.89 \pm 0.17 \pm 0.29} $ \\
\decxCC &  $ \rm { 9.91 \pm 0.31 \pm 0.27} $ \\ \hline
 \end{tabular}
% -------------------------------------------------------------------
\caption{\label{br_ratio_results} Branching ratio results.
The first error is statistical and the second systematic.}
\end{center}
\end{table}
The measurements of these branching ratios are correlated.
The correlation coefficients between branching ratios
calculated using the statistical errors on each branching ratio
are given in Table~\ref{correl_coeff}.
\begin{table}
\begin{center}
\begin{tabular}{ l c c } \hline
Sample   & \decxB\     & \decxCC\      \\ \hline
\decxA\  & \CorrelOne\ & \CorrelTwo\   \\
\decxB\  &            & \CorrelThree\  \\ \hline
\end{tabular}
\caption{\label{correl_coeff} The correlation coefficients between each measurement.}
\end{center}
\end{table}

%
% ......... Systematics ................................
%
\section{Systematic errors}
\label{syst_err}

The systematic errors on the branching ratios
are shown in Table~\ref{table_syst_err}.
\begin{table}[t]
\begin{center}
% -------------------------------------------------------------------
% {\input{./inputs/SysError3.tex}} Nov 24 version
 \begin{tabular}{| l | c | c | c |} \hline
   & \multicolumn{3}{c|}
 {Systematic error for each selection (\%)} \\ \cline{2-4}
  & { \decxA  }  & { \decxB  }  & { \decxCC } \\ \hline 
            MC statistics & $ 0.11 $ & $ 0.17 $ & $ 0.14 $ \\
          $a_1$ modelling & $ 0.01 $ & $ 0.13 $ & $ 0.15 $ \\
 {1-cluster \pz\ threshold} & 
$ 0.03 $ 
  & 
$ 0.02 $ 
  & 
$ 0.03 $ 
  \\ 
 {2-cluster \pz\ threshold} & 
$ 0.03 $ 
  & 
$ 0.11 $ 
  & 
$ 0.11 $ 
  \\ 
 {Anode plane cut} & 
$ 0.00 $ 
  & 
$ 0.05 $ 
  & 
$ 0.03 $ 
  \\ 
 {Photon conversions }& 
$ 0.06 $ 
  & 
$ 0.03 $ 
  & 
$ 0.02 $ 
  \\ 
 {Energy smearing }& 
$ 0.01 $ 
  & 
$ 0.01 $ 
  & 
$ 0.03 $ 
  \\ 
 {Radiative clusters }& 
$ 0.03 $ 
  & 
$ 0.00 $ 
  & 
$ 0.00 $ 
  \\ 
 {Energy scale }& 
$ 0.03 $ 
  & 
$ 0.03 $ 
  & 
$ 0.04 $ 
  \\ 
    Signal BR uncertainty & $ 0.01 $ & $ 0.05 $ & $ 0.03 $ \\
            Bias factors  & $ 0.02 $ & $ 0.03 $ & $ 0.02 $ \\
      Unmodelled channels & $ 0.00 $ & $ 0.02 $ & $ 0.04 $ \\
 \hline
   Non-$\tau$ backgrounds & $ 0.02 $ & $ 0.04 $ & $ 0.01 $ \\
                  \tauE\  & $ 0.04 $ & $ 0.07 $ & $ 0.08 $ \\
                  \tauM\  & $ 0.05 $ & $ 0.08 $ & $ 0.01 $ \\
                 \decxKZ\ & $ 0.06 $ & $ 0.07 $ & $ 0.04 $ \\
        Other backgrounds & $ 0.00 $ & $ 0.03 $ & $ 0.03 $ \\
 \hline \hline
                    Total & $ 0.16 $ & $ 0.29 $ & $ 0.27 $ \\
 \hline
 \end{tabular}
% -------------------------------------------------------------------
\caption{\label{table_syst_err} Systematic errors.}
\end{center}
\end{table}
The efficiencies for detecting the signals and backgrounds are taken from
\MC\ and special attention is paid to the \MC\ modelling.
The first half of Table~\ref{table_syst_err} gives the systematic errors due
to aspects of the analysis such as \MC\ modelling of hadronic showers, 
track finding, conversion finding, radiative photons and energy scale.
The second half of Table~\ref{table_syst_err} gives the systematic 
errors due to the backgrounds.
We briefly describe the individual contributions to the systematic errors.

The $ a_1 \rightarrow  \pi^- 2\pz $ decay was simulated using the IMR 
model, as described in section~\ref{ev_sel}.
The IMR model gives a better description of the 
$ a_1^- \rightarrow \pi^- \pi^+ \pi^- $ data
primarily because of the inclusion of the polynomial 
background term~\cite{imr_model}.
The uncertainty in the contribution of the polynomial background
was measured in \cite{opal_had_paper}
to be approximately $ 17 \% $.
The systematic error in the branching ratios due to the uncertainty in the
$ a_1 $ model is conservatively estimated by varying the normalization of the polynomial
background by $ \pm 50 \% $ ($ 3 \sigma$).

The \MC\ models the hadronic showering in $\tau$ decays reasonably well.
The thresholds for the neutral clusters, one-cluster and two-cluster 
\pzs, as well as other \reqmnts, have been carefully studied
to avoid any poorly modelled regions due to low energy hadronic clusters.
However, as the energy thresholds are lowered, deviations between
the data and \MC\ appear since clusters close to the track 
from hadronic interactions are accepted into the $\pi^0$-finding algorithm.
The lower energy thresholds of both the one and two cluster \pz\ cases
are varied to estimate the systematic error due to 
the \MC\ modelling of energy deposition in the electromagnetic
calorimeter.
The lower energy threshold of the one-cluster \pz\ case is varied 
from $2.0$ to $3.0$~GeV and the energy threshold of the 
two-cluster \pz\ case is varied from $2.6$ to $4.0$~GeV.
The maximum variation from each nominal branching ratio
is taken as the systematic error in each case.

The effect of tracks passing close to the anode plane of the OPAL jet
chamber is considered as a source of systematic error.
The \MC\ does not perfectly model the position of
tracks close to the anode plane.
The branching ratios were recalculated by removing
tracks within  $0.25^\circ$ of the anode plane.
The resulting change in the branching ratios is taken as the systematic
error.

Approximately $5 \%$ of the identified \pzs\ are composed of a 
neutral cluster and a photon conversion.
Although the \MC\ does a reasonable job of modelling conversions, 
there are minor discrepancies between data and \MC.
The ratio of data to \MC\ jets with an identified conversion pair
is found to be $0.985 \pm 0.013$.
The branching ratios are calculated with \MC\ jets containing a conversion
weighted to simulate a $\pm 3\sigma$ ($\pm 4 \%$) 
change in conversion identification efficiency.
The maximum variation from each nominal branching ratio
is taken as the systematic error. 
The sensitivity of the results to photon conversions is also checked by 
dropping the conversion routine and recalculating the branching ratios.

The energy of each electromagnetic calorimeter cluster in the \MC\ 
is  smeared. 
The uncertainty due to this smearing is assessed
by varying the amount of smearing applied by $\pm 20 \%$.
The change in the branching ratios is taken as the systematic error.

Clusters were considered to be due to radiative photons if
the mass of the track and cluster was greater than $3$~GeV.
These clusters are ignored by the $\pi^0$-algorithm.
This requirement was removed and the change in the branching ratios
was taken as the systematic error.

The energy scale uncertainty reflects
the uncertainty in the electromagnetic calorimeter calibration 
between \MC\ and data.
To determine the systematic uncertainty the clusters
are rescaled by $ \pm 0.2 \% $
and the largest effect on the 
branching ratios is taken as the systematic uncertainty.

The selection efficiencies depend on the relative 
branching ratios of the individual decay modes.
The signal branching ratios were calculated using the PDG
branching ratios and uncertainties~\cite{PDG96}.
The dependence of the efficiency and the decay branching ratios
on the signal branching ratios was evaluated and included as a systematic
uncertainty.

A small correction must be applied to the branching ratios to 
correct the slight bias introduced by
the \taupair\ selection criteria.
The dependence of the bias factor on the \taupair\ selection and
on the branching ratios of each channel is found to be relatively small.
In addition, other \MC\ samples using a different electromagnetic shower
model give similar results.
The systematic error on each branching ratio is 
calculated directly using the bias factor error.

The \MC\ used in this analysis did not include some decay modes
defined as signals by the PDG~\cite{PDG96}.
Table~\ref{sig_chan} shows these decay modes.
The systematic effect of the \decxE\ mode on the \decxB\ and \decxCC\ samples is
calculated using the PDG branching ratio and the assumption that 
the efficiency for \decxE\ is the same as that for \decxD.
The uncertainty for each of these modes is assumed to be $50 \%$
of their respective branching ratios.
The total systematic error is calculated from these uncertainties.

The systematic errors on the branching ratios due to non-$\tau$ 
and $\tau$ backgrounds are calculated from the uncertainties given in
Tables~\ref{non_tau_bkgrd} and \ref{corr_backgrd}, respectively.
An additional error has been added to the \decxKZ\ decays to 
account for any uncertainty in the energy deposited by a neutral
hadron in the electromagnetic calorimeter.
A 10\% error was added to the \decxKZ\ background in the \decxA\ sample.
A second error was added to the \decxKZ\ background in all three samples to
account for possible migration of the background from one \pz\
sample to another.

%
% ......... Discussion  ................................
%
\section{Discussion}
\label{discussion}

The \decxA\ branching ratio is measured to be \reszerox\ and is compared with
other published results in figure~\ref{br_comp_0}.
The \decxA\ branching ratio agrees well with the PDG world average of 
$(11.70 \pm 0.11)\%$~\cite{PDG96} but is slightly above
the recent CLEO measurement of $(11.52 \pm 0.13)\%$~\cite{CLEO96}
and the theoretical prediction of
$(11.65 \pm 0.06)\% $ made by Decker and Finkemeier~\cite{decker_finkemeier}.

The \decxA\ branching ratio can be used to measure the ratio of the
charged current coupling constants of taus and muons.
Lepton universality requires that the weak charged current gauge coupling
strengths be identical: $ g_e = g_\mu = g_\tau $.
The $ g_\mu / g_\tau $ ratio is probed by comparing the decays
$h^- \rightarrow \mu^- \overline{\nu}_\mu$ and \decxA.
An expression for $g_\tau^2 / g_\mu^2$ is given by
\bdm
{ \frac{g_\tau^2}{g_\mu^2} } = 
\left \{ \frac{2  m^2_\mu }{m^2_\tau} \right \}
\frac{ {\rm BR}(\decxA) }
     { H_{\pi} + H_{\rm K}}  
\, ,
\edm
where 
\bdm
H_h = 
( 1 + \delta_h )
\left ( \frac{ \tau_{\tau} m_{\tau} }{ \tau_h m_h } \right )
\left [ \frac{1 - \left ( m_h/m_\tau \right )^2 }
             {1 - \left ( m_\mu/m_h \right )^2 } \right ]^2
{\rm BR} \, (h^- \rightarrow \mu^- \overline{\nu}_\mu)
\, ,
\edm
and $\delta_h$ is an 
electromagnetic radiative correction~\cite{decker_finkemeier}.
The $\pi^- \rightarrow \mu^- \overline{\nu}_\mu$ and
${\rm K^-} \rightarrow \mu^- \overline{\nu}_\mu$ branching ratios, 
the masses of the tau, muon and the pion, and the lifetime of 
the pion are taken from the PDG~\cite{PDG96}.
The ratio $g_\tau / g_\mu $ is found to be
\bdm
{ \frac{g_\tau}{g_\mu} } = \UnivResult \pm \UnivResultErr
\, ,
\edm
where the error on $g_\tau / g_\mu $ is dominated by the uncertainties in the
\decxA\ branching ratio and the OPAL $\tau$ lifetime~\cite{opal_tau_lifetime}.
The point in figure~\ref{univ_1} shows the result from this work plotted
against the $\tau$ lifetime.
The Standard Model prediction is shown 
as the shaded band with a width reflecting the uncertainty in $m_\tau$.
If we recalculate the world average \decxA\ branching ratio including
our result and use the PDG value for the tau lifetime~\cite{PDG96}, 
then the ratio $g_\tau / g_\mu$ is found to be $\NewUnivRes \pm \NewUnivResErr$.

The ratio of $ g_\tau / g_\mu $ can also be measured by comparing the \tauE\ 
and $\rm \mu \rightarrow e^- \overline{\nu}_e \nu_\mu$ decays.
Using the PDG values for the \tauE\ and $\tau$ lifetime gives
$ g_\tau / g_\mu = 1.000 \pm 0.003 $.
Although this result is more precise, the two measurements are
complementary.
The measurement using the \decxAP\ and 
$\pi^- \rightarrow \mu^- \overline{\nu}_\mu$ decays probes the couplings to 
a longitudinal W boson, while the measurement using the
\tauE\ and $\rm \mu \rightarrow e^- \overline{\nu}_e \nu_\mu$ decays probes the 
couplings to a transverse W.

%
% .....RPV section

The decay \thn\ is also sensitive to the presence of physics
beyond the minimal Standard Model.
As an example, in $R$-parity violating (RPV) 
extensions of supersymmetric models~\cite{RPVth}
\thn\ and \hmn\ receive an additional contribution from
the exchange of a right-handed scalar d-type quark, \sdkR ,
where $k$ is a family index.
This exchange can be cast in a $V-A \otimes V-A$ form
leading to a modification of the Standard Model predicted couplings
by a term proportional to
$|\lpRPV{ijk}|^2/\msdkRsq $
as outlined in~\cite{BGH}.
The RPV Yukawa couplings probed by \thn\ and \hmn\ decays are
\lpRPV{31k} and \lpRPV{21k} respectively.
Note that, 
under the assumption that only one RPV Yukawa coupling is non-zero,
the decays \tKn\ and \Kmn\ are unaffected by RPV
since they involve products of two different coupling constants.

Using the formalism developed in~\cite{BGH},
limits can be set on \lpRPV{31k} and \lpRPV{21k}
from the expression,
derived here for the first time,
\begin{displaymath}
  \frac{|\lpRPV{31k}|^2 - |\lpRPV{21k}|^2}{\msdkRsq } =
  2 \sqrt{2} G_{\mathrm{F}} V_{\mathrm{ud}}
  \left[ \frac{2 \mmu ^2}{\mtau ^2}
           \frac{\mathrm{BR}(\thn )}{H_{\pi}}
         - 1 - \frac{H_K}{H_{\pi}}
  \right].
\end{displaymath}
Making the assumption that only one Yukawa coupling is non-zero
leads to the following 95\% confidence level limits,
calculated using \msdkR\ = 100 GeV, of
\[  
|\lpRPV{31k}| < \LsssOPthree 
\;\;\; \mbox{\rm and} \;\;\; 
|\lpRPV{21k}| < \LsssOPtwo
\]
using the OPAL \decxA\ and tau lifetime~\cite{opal_tau_lifetime} measurements and 
\[ 
|\lpRPV{31k}| < \LsssWAthree
\;\;\; \mbox{\rm and} \;\;\; 
|\lpRPV{21k}| < \LsssWAtwo
\]
using the world average \thn\ branching ratio and tau lifetime.
Previous measurements have quoted the  68\% ($1\sigma$) confidence limits
\cite{tpnlim}.
Using the OPAL measurements, the 68\% ($1\sigma$) confidence limits are
$|\lpRPV{31k}| < \LsOPthree$ 
and $|\lpRPV{21k}| < \LsOPtwo$, 
while  the
world average results give 68\% confidence limits of
$|\lpRPV{31k}| < \LsWAthree$ 
and $|\lpRPV{21k}| < \LsWAtwo$.
The limit on \lpRPV{21k} set using this method is competitive
with the present best limit derived from pion decay 
to electrons and muons~\cite{BGH}.
Limits on \lpRPV{31k} have been obtained previously using
the decay \tpn\ ~\cite{tpnlim}.
Our new calculation using \thn\ has several advantages over this method.
Firstly, BR(\thn ) is more precisely known than BR(\tpn ).
Additionally, BR(\thn ) is a directly measured quantity
making the limits  more experimentally compelling
than those derived from \tpn.

The \decxB\ branching ratio is measured to be \resonex\ and is 
compared to other published branching ratios in figure~\ref{br_comp_0}.
Some of the branching ratios shown have been corrected by the PDG in order to
treat the kaon backgrounds in a uniform manner.
The \decxB\ branching ratio result measured in this work agrees well with the 
PDG average ($ 25.76 \pm  0.15 \% $) and the previous measurements.

The Conserved Vector Current hypothesis~\cite{cvc_hyp_1} can be used
to predict the \decxBP\ branching ratio from low energy
$\epm \rightarrow \, \pi^+ \pi^- $ data.
A number of predictions for the \decxBP\ branching ratio have been 
made~\cite{ks_model,tsai}.
A recent review by Eidelman and Ivanchenko~\cite{cvc_pred} predicted the
branching ratio to be $(24.25 \pm 0.77) \%$.
The \decxB\ branching ratio measured in this work
should be modified by subtracting the \decxBK\ branching
ratio~\cite{PDG96} giving a \decxBP\ branching ratio of $(25.73 \pm 0.31)\%$.
The result measured here is consistent (within two standard deviations)
with the CVC prediction.

The \decxCC\ branching ratio is measured to be \restwox,
in comparison to the PDG average for the \decxCC\ branching ratio 
of $ (10.48 \pm 0.35) \% $ \cite{PDG96}.
The PDG number quoted is the sum of the average values for the
branching ratios of the tau decay to the $h^-2\pi^0\nu_\tau$,
$h^-3\pi^0\nu_\tau$, $h^-4\pi^0\nu_\tau$, $\pi^-\overline{K}^0\nu_\tau$
and $K^-\overline{K}^0\nu_\tau$ modes.
%
% ......... Conclusions  ................................
% 
\section{Conclusions}

The branching ratios of \decxA, \decxB\ and \decxCC\
decays have been measured with the OPAL detector at LEP.
The branching ratios are
\begin{center}
\begin{tabular}{l c c c } \\
BR(\decxA)  & = & \reszerox &  \\
BR(\decxB)  & = & \resonex &   \\
BR(\decxCC) & = & \restwox & , \\
\end{tabular}
\end{center}
where the first error is statistical and the second error is systematic.
These new measurements are more precise than previous OPAL 
measurements and supersede those results.

The \decxA\ branching ratio measured in this work 
is found to be in good agreement with previous measurements.
The ratio of the charged current coupling constants of 
muons and taus using the 
\decxA\ branching ratio is found to be 
$g_\mu / g_\tau = ( \UnivResult \pm \UnivResultErr ) $,
consistent with lepton universality.
The \decxA\ branching ratio is used to place
limits on supersymmetric $R$-parity violating Yukawa couplings.
The \decxB\ branching ratio found in this work 
is in good agreement with the previous results 
and also with the Conserved Vector Current prediction.
Finally, the \decxCC\ branching ratio measured in this work is found
to be consistent with the current PDG world average.
%
% ......... Acknowledgements .............................
%
\bigskip\bigskip\bigskip\bigskip
\appendix
\par
\noindent
{\large\bf Acknowledgements}\\
\noindent
We particularly wish to thank the SL Division for the efficient operation
of the LEP accelerator at all energies
 and for
their continuing close cooperation with
our experimental group.  We thank our colleagues from CEA, DAPNIA/SPP,
CE-Saclay for their efforts over the years on the time-of-flight and trigger
systems which we continue to use.  In addition to the support staff at our own
institutions we are pleased to acknowledge the  \\
Department of Energy, USA, \\
National Science Foundation, USA, \\
Particle Physics and Astronomy Research Council, UK, \\
Natural Sciences and Engineering Research Council, Canada, \\
Israel Science Foundation, administered by the Israel
Academy of Science and Humanities, \\
Minerva Gesellschaft, \\
Benoziyo Center for High Energy Physics,\\
Japanese Ministry of Education, Science and Culture (the
Monbusho) and a grant under the Monbusho International
Science Research Program,\\
German Israeli Bi-national Science Foundation (GIF), \\
Bundesministerium f\"ur Bildung, Wissenschaft,
Forschung und Technologie, Germany, \\
National Research Council of Canada, \\
Research Corporation, USA,\\
Hungarian Foundation for Scientific Research, OTKA T-016660, 
T023793 and OTKA F-023259.\\

%
% ......... Bibliography ................................
%

% .......... Figure 1: number of clusters
\newpage
\bfi
\begin{center}
\mbox{\epsfig{file=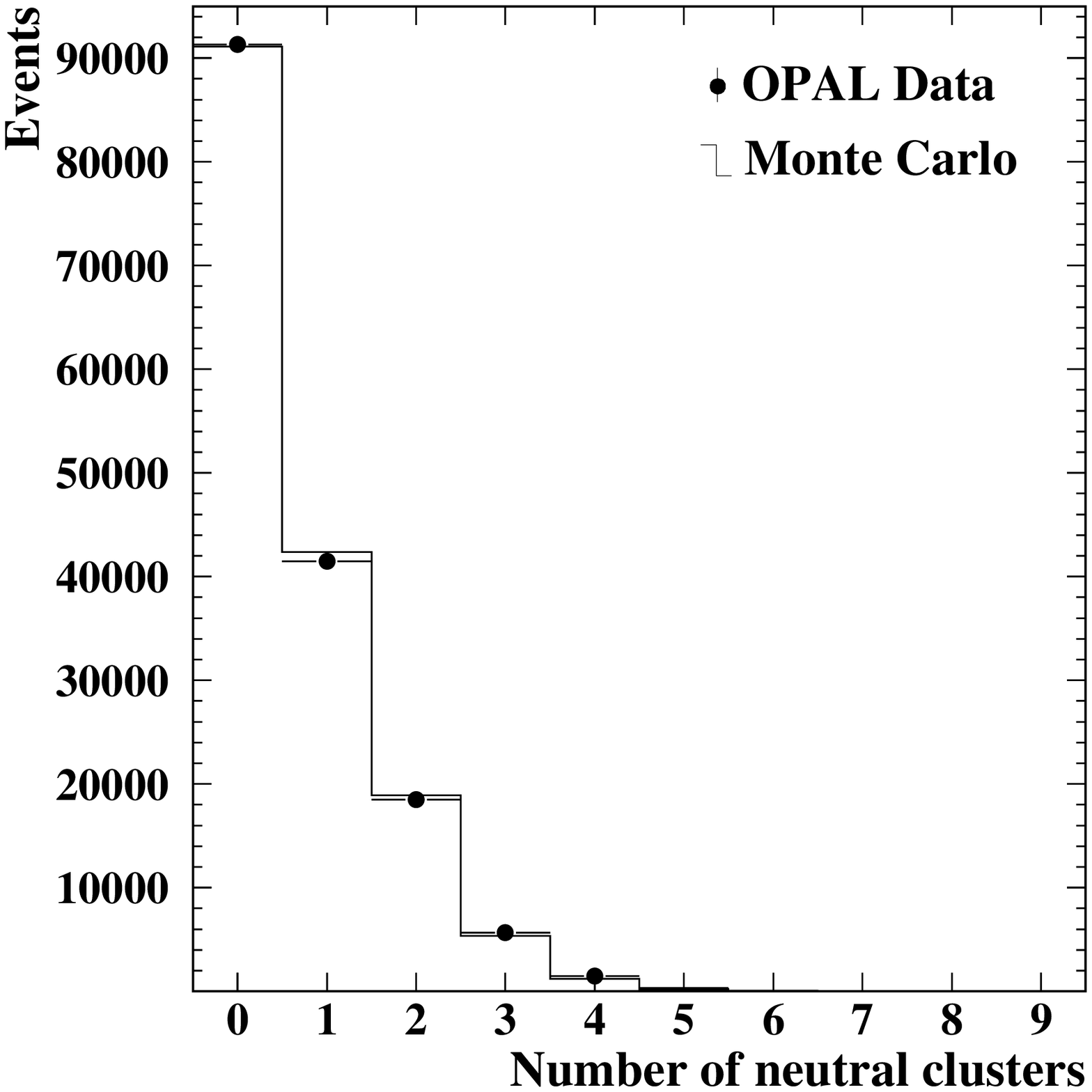,width=8.5cm,bbllx=0pt,bblly=0pt,bburx=567pt,bbury=567pt}}
\mbox{\epsfig{file=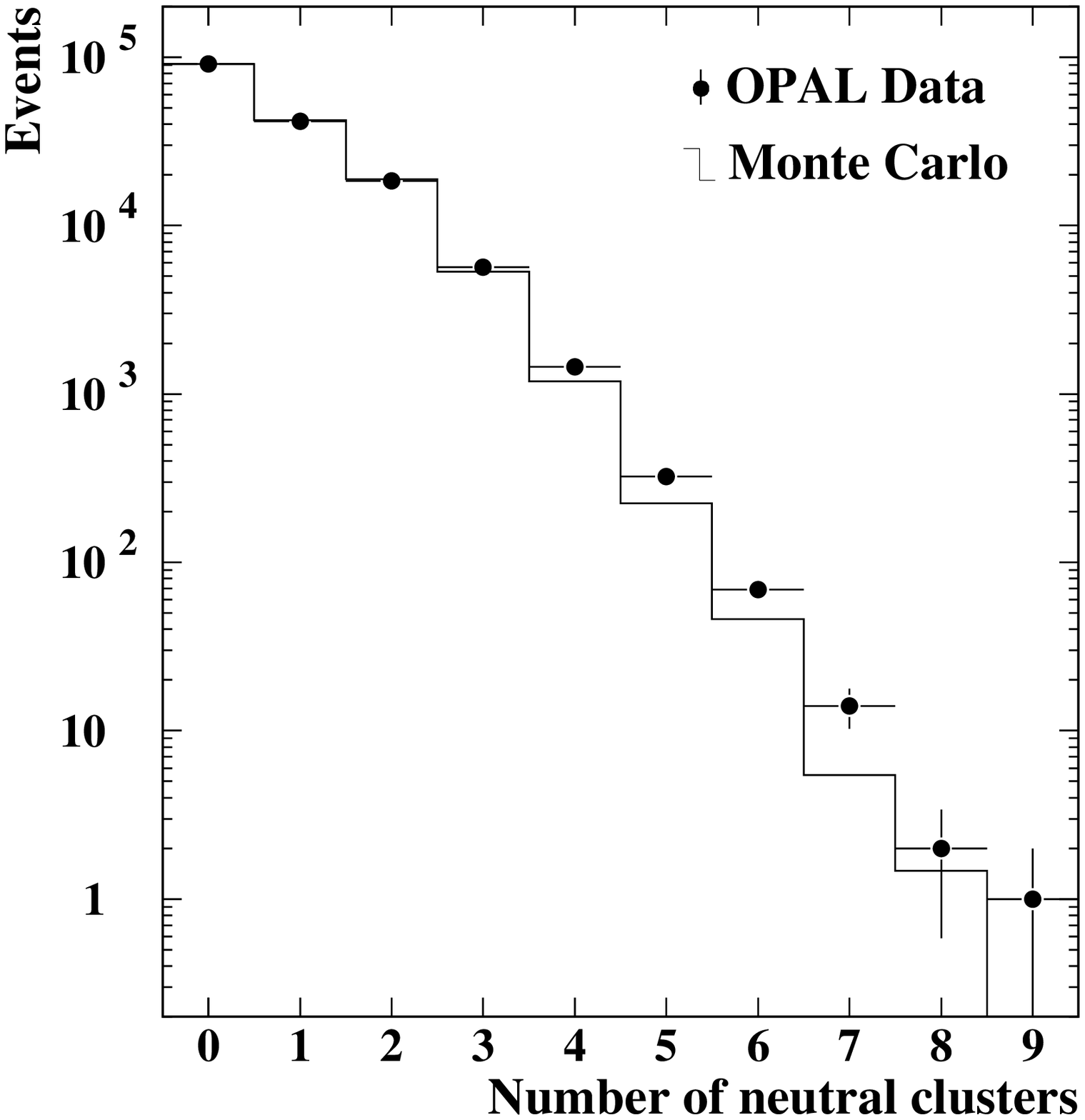,width=8.5cm,bbllx=0pt,bblly=0pt,bburx=567pt,bbury=567pt}}
\caption{\label{nnclus2} The distribution of the  number of neutral clusters 
per jet for one-prong selected jets.
The distribution is shown in both a linear vertical scale (left plot)
and a logarithmic vertical scale (right plot).}
\end{center}
\efi

% ........... Figure 2: radiative cluster energy
\bfi
\begin{center}
\mbox{\epsfig{file=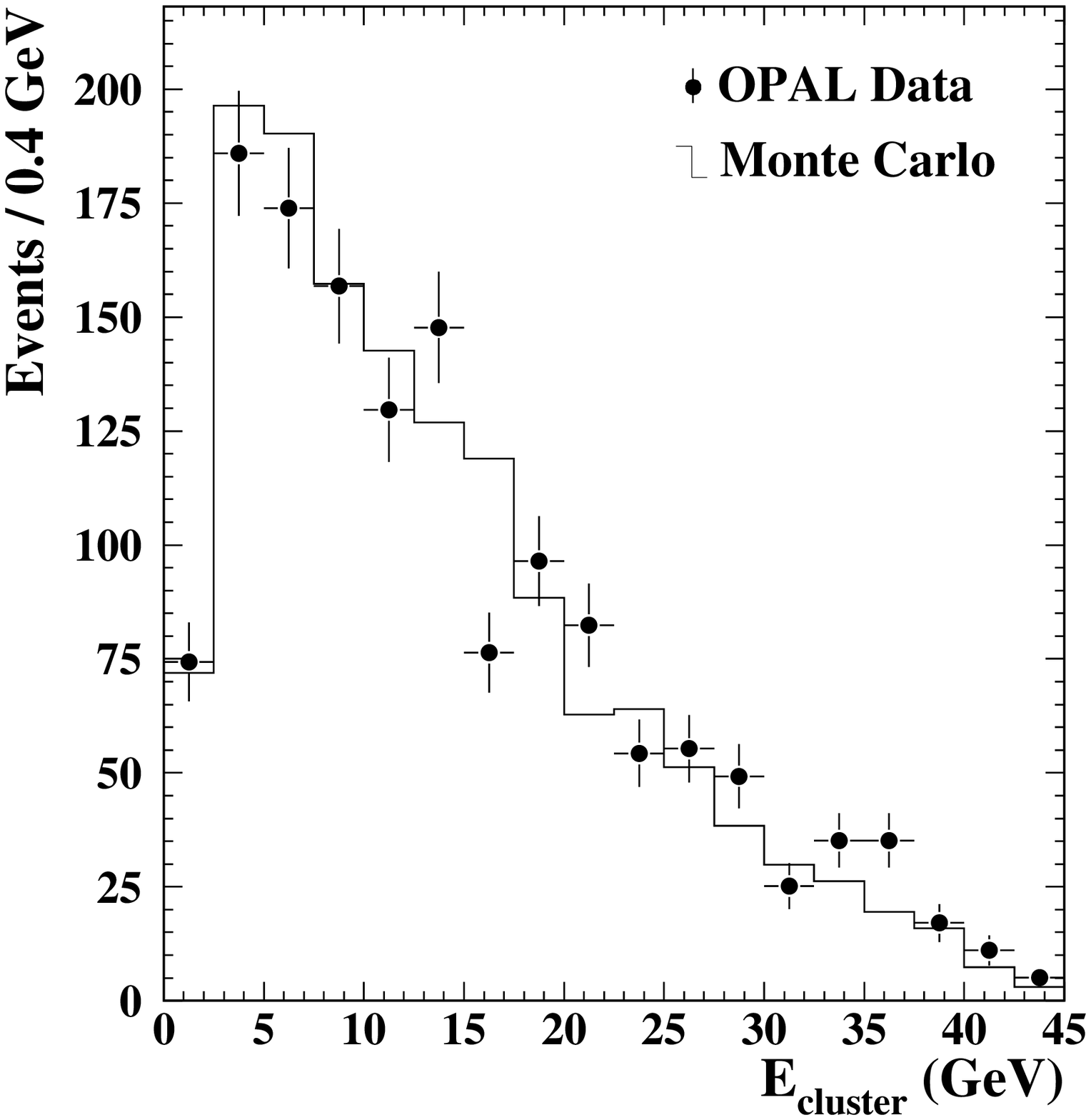,width=8.5cm,bbllx=0pt,bblly=0pt,bburx=567pt,bbury=567pt}}
\caption{\label{rad_ener} The distribution of the energy of 
clusters identified as radiative photons.}
\end{center}
\efi

% ........... Figure 3: two cluster mass
\bfi[t]
\begin{center}
\mbox{\epsfig{file=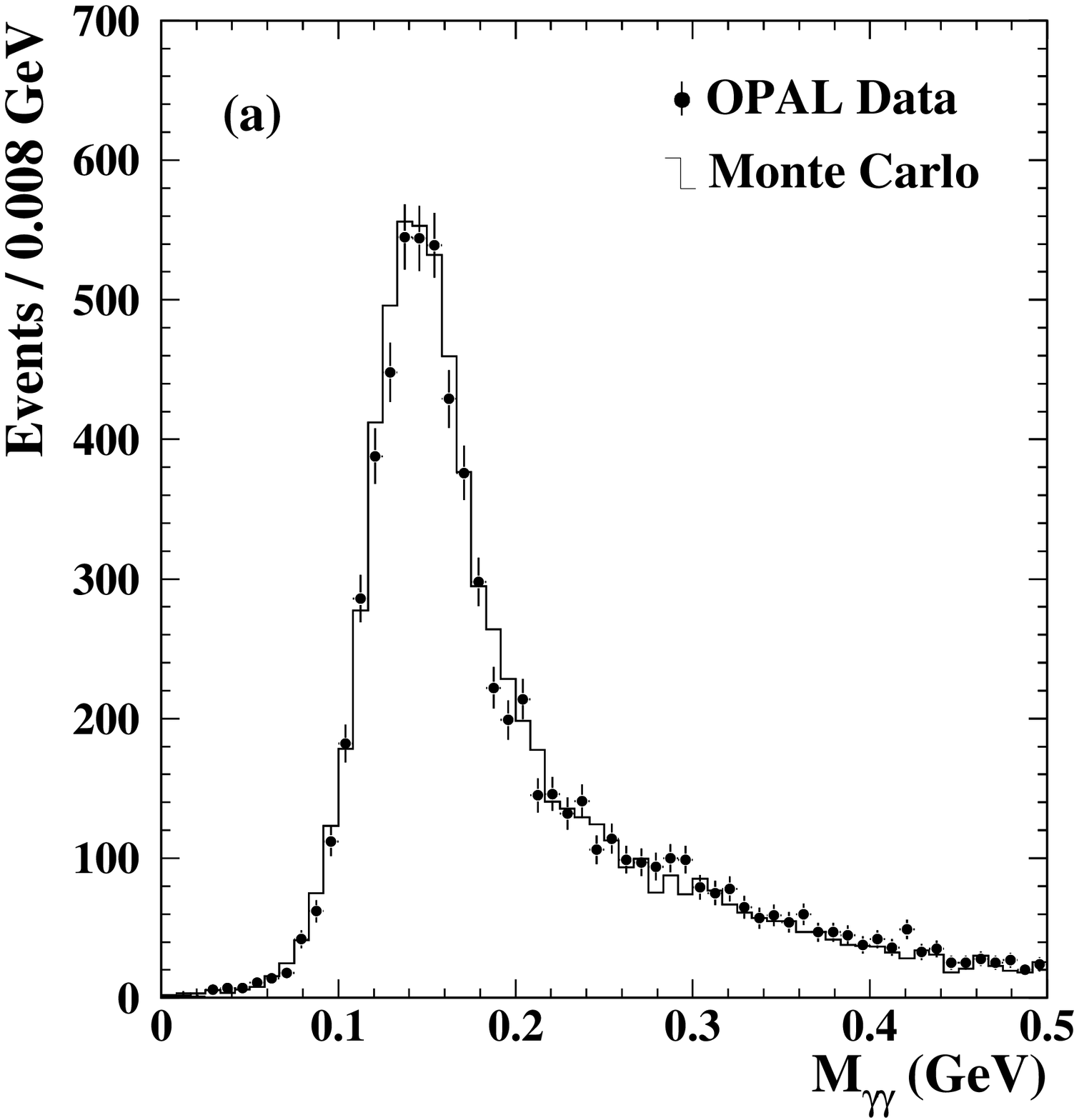,width=8.75cm,bbllx=0pt,bblly=0pt,bburx=567pt,bbury=567pt}}
\mbox{\epsfig{file=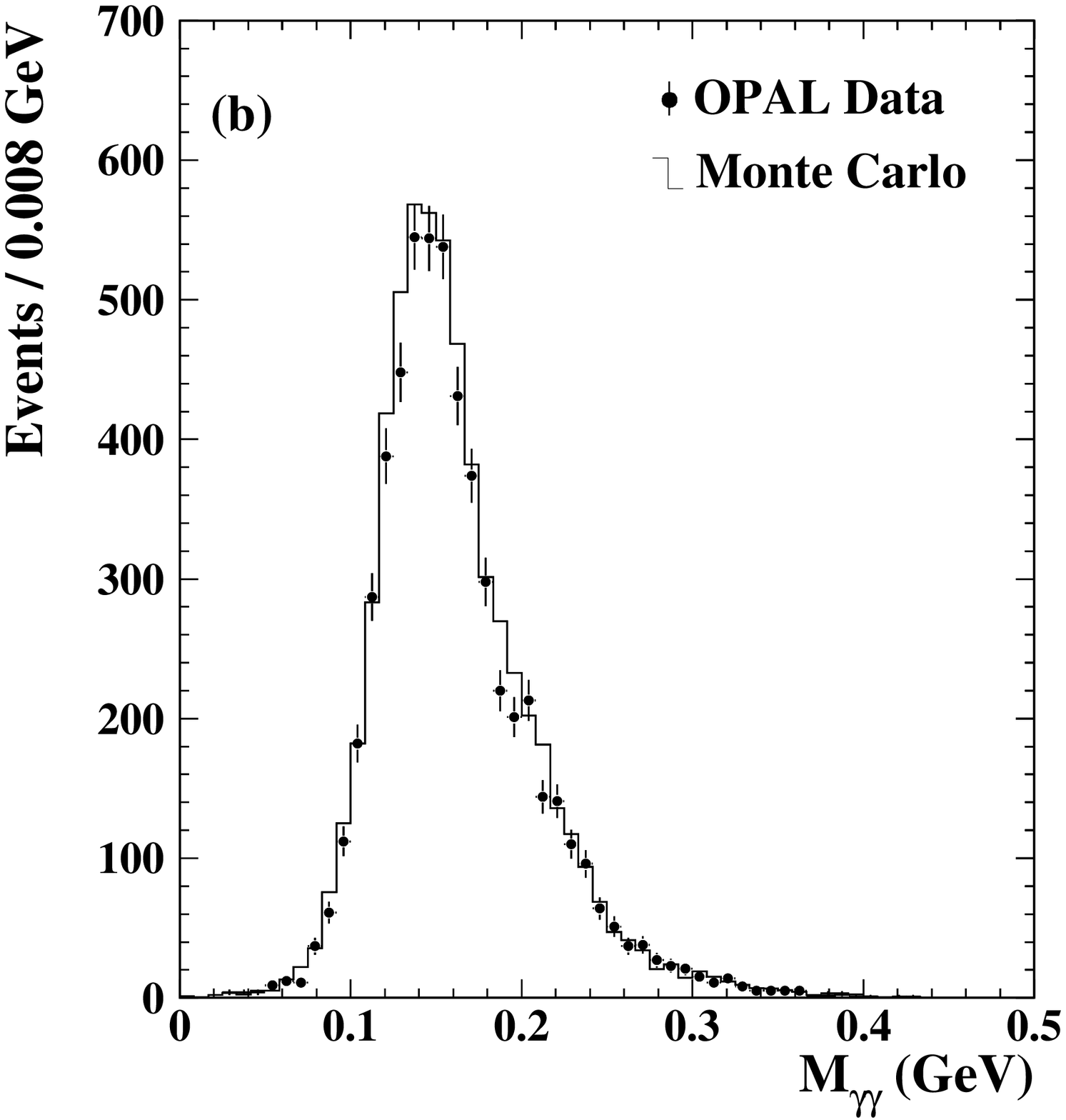,width=8.75cm,bbllx=0pt,bblly=0pt,bburx=567pt,bbury=567pt}}
\caption{\label{mass_gg} 
(a) The invariant mass distribution of two neutral clusters in data and \MC\ 
for jets with two neutral clusters with energy between $0.5$ and $9.0$~GeV.
(b) The invariant mass distribution for the neutral cluster pairs selected
by the \pz\ algorithm.}
\end{center}
\efi

% ........... Figure 4: Number and type of pizeros
\bfi
\begin{center}
\mbox{\epsfig{file=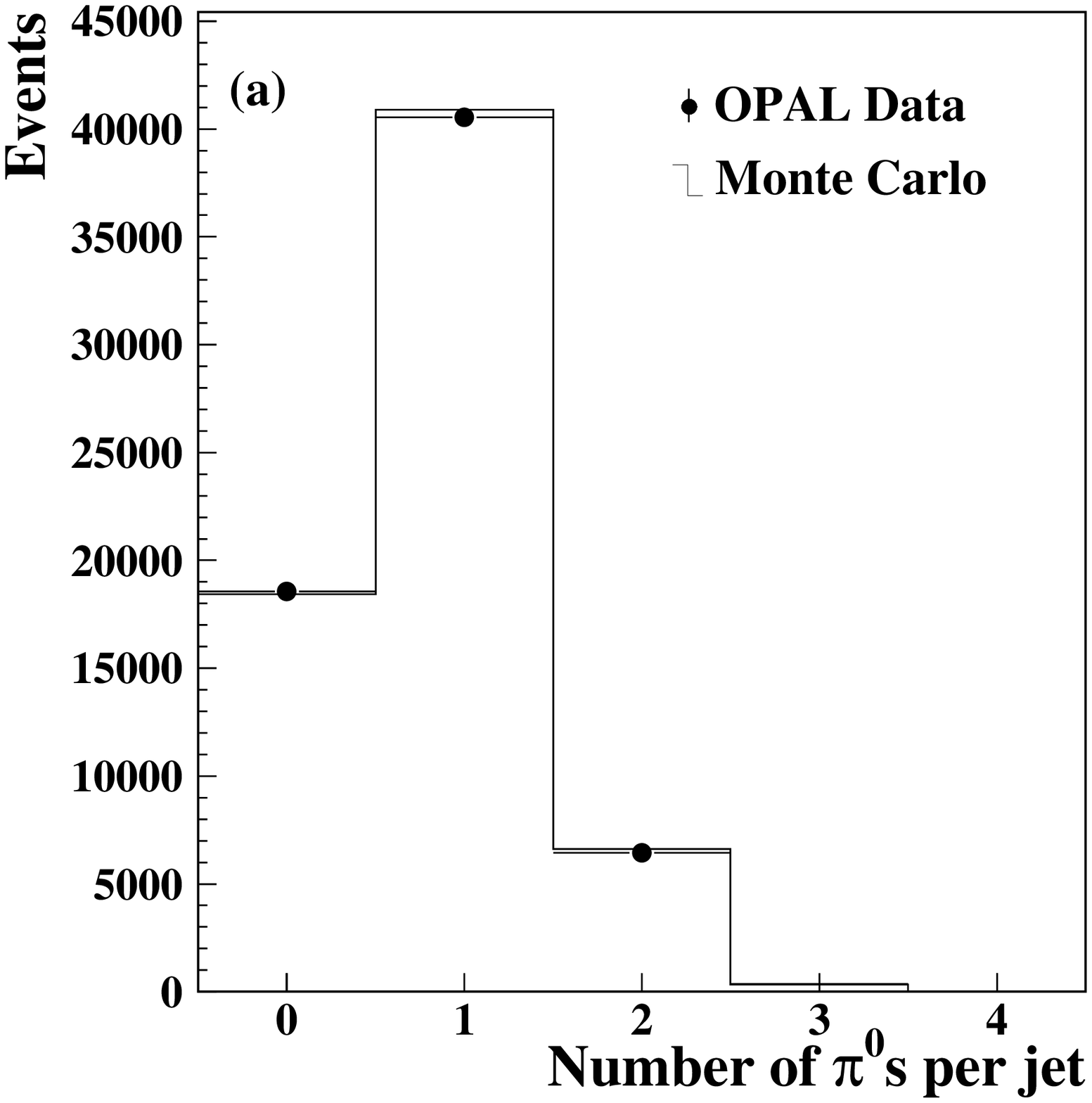,width=8.75cm,bbllx=0pt,bblly=0pt,bburx=567pt,bbury=567pt}}
\mbox{\epsfig{file=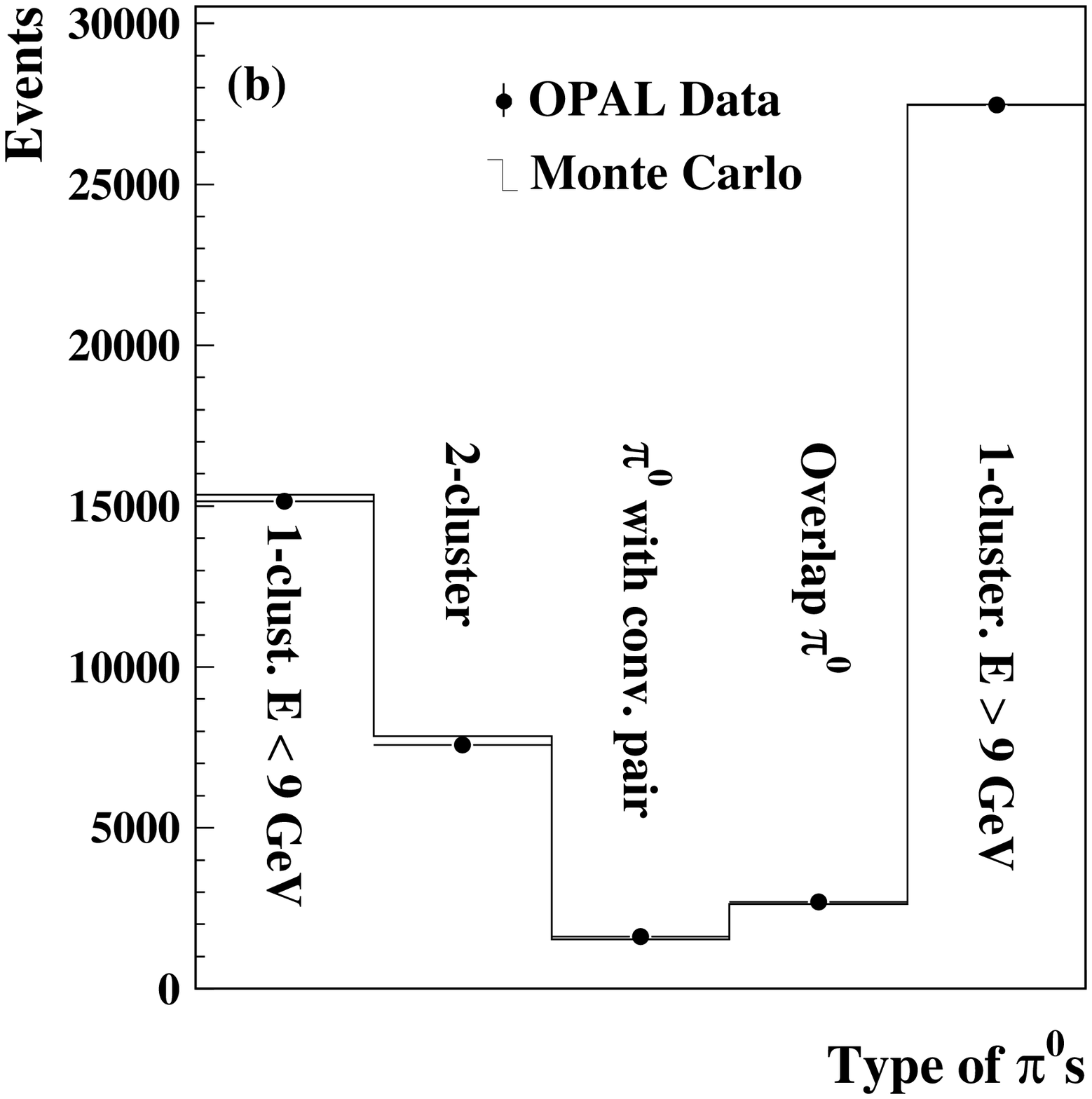,width=8.75cm,bbllx=0pt,bblly=0pt,bburx=567pt,bbury=567pt}}
\caption{\label{num_pi0} 
(a) The distribution of the number of \pzs\ in each one-prong jet identified 
by the \pz\ finding algorithm. 
(b) The distribution of the types of \pzs\ identified in one-prong jets.}
\end{center}
\efi

% ........... Figure 5: Energy of pizeros
\bfi[t]
\begin{center}
\mbox{\epsfig{file=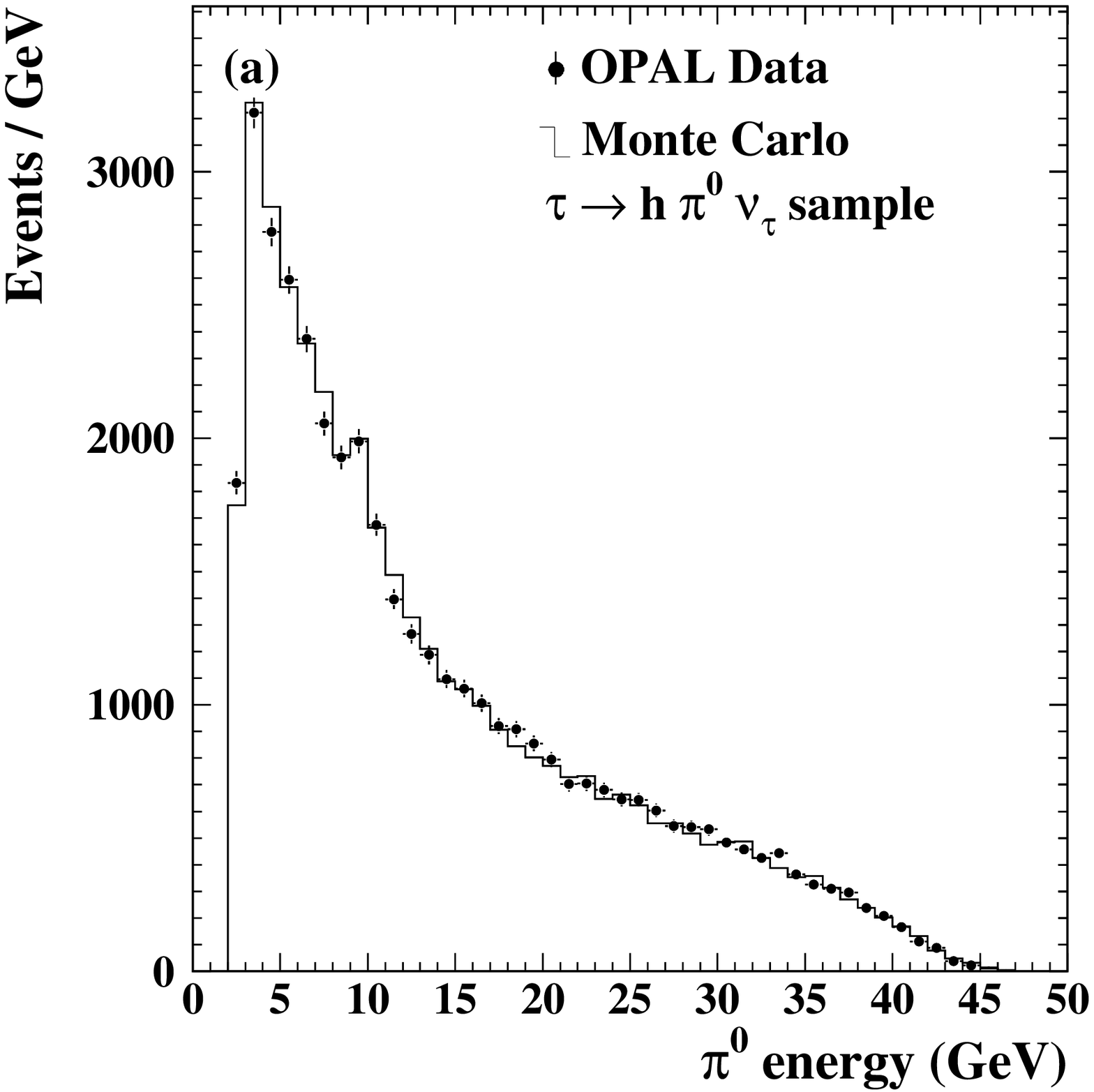,width=8.75cm,bbllx=0pt,bblly=0pt,bburx=567pt,bbury=567pt}}
\mbox{\epsfig{file=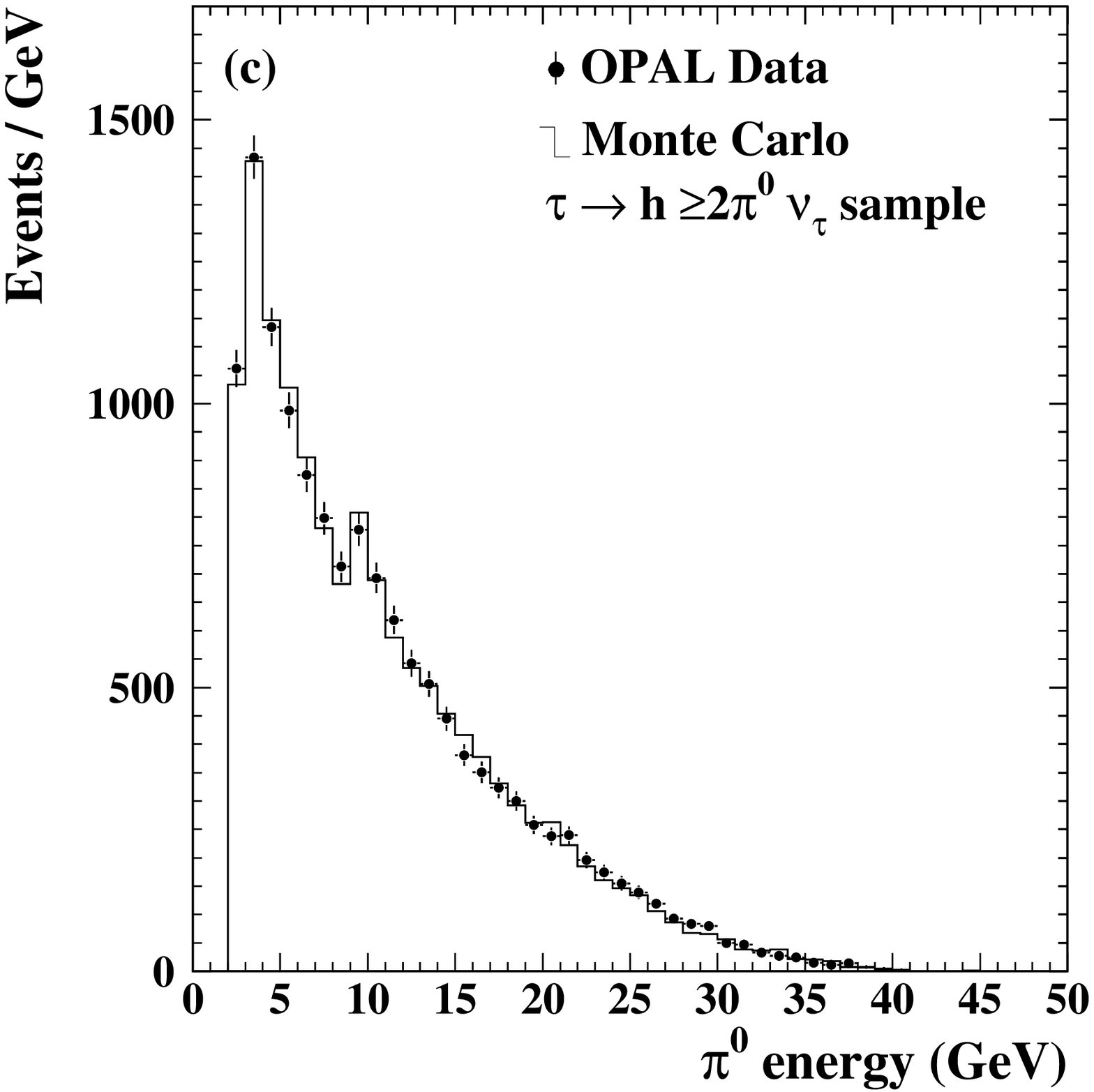,width=8.75cm,bbllx=0pt,bblly=0pt,bburx=567pt,bbury=567pt}}
\mbox{\epsfig{file=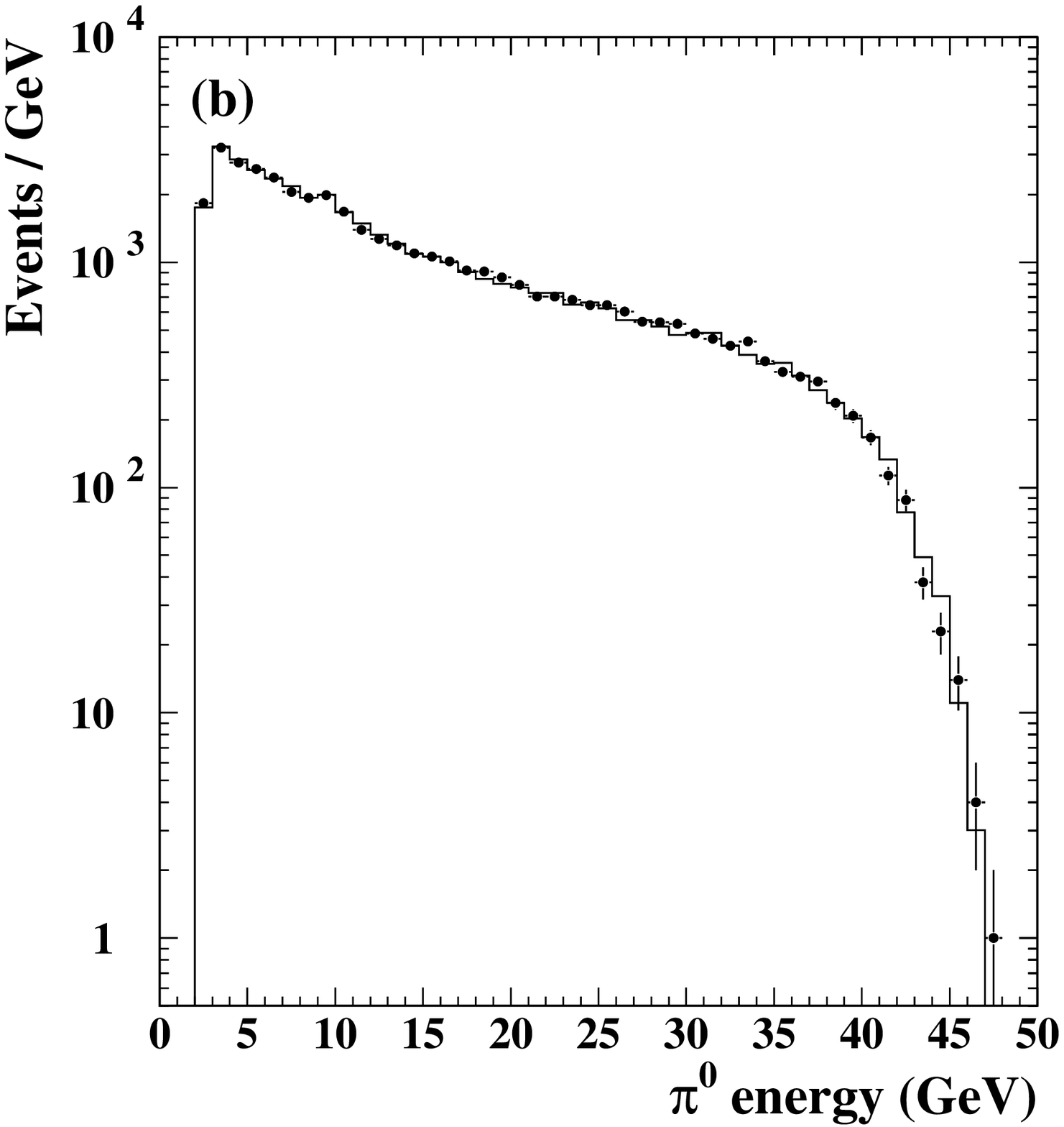,width=8.75cm,bbllx=0pt,bblly=0pt,bburx=567pt,bbury=567pt}}
\mbox{\epsfig{file=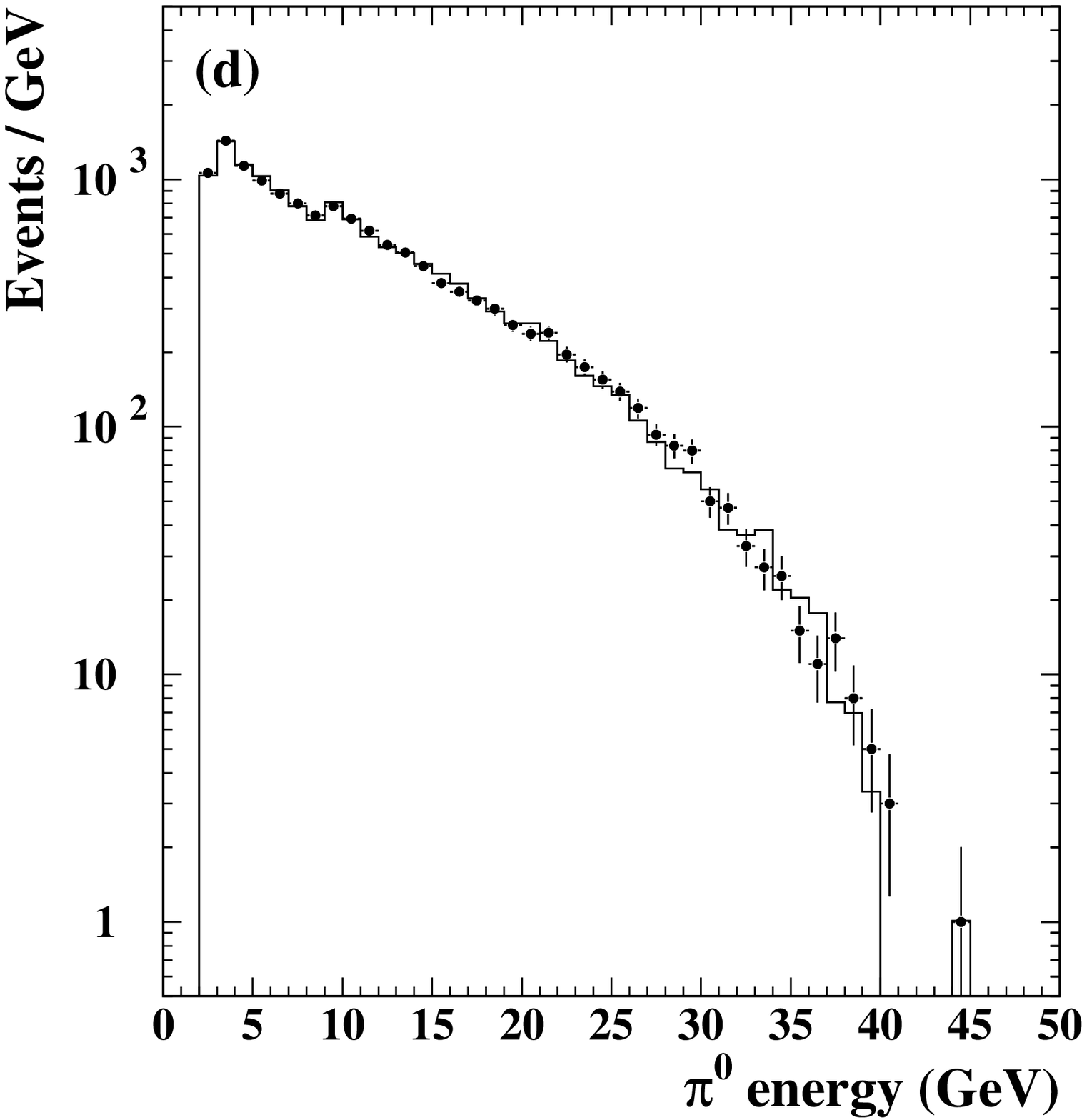,width=8.75cm,bbllx=0pt,bblly=0pt,bburx=567pt,bbury=567pt}}
\caption{\label{en_pi0_x} 
The energy distribution of reconstructed \pzs\ shown in both linear and logarithmic scales.
(a) and (b) show the energy of \pzs\ in \decxB\ jets.
(c) and (d) show the energy of \pzs\ in \decxCC\ jets.
The structure observed at 9 GeV corresponds to the one-cluster \pz\ threshold.
 }
\end{center}
\efi

% ........... Figure 6: jet mass
\bfi[t]
\begin{center}
\mbox{\epsfig{file=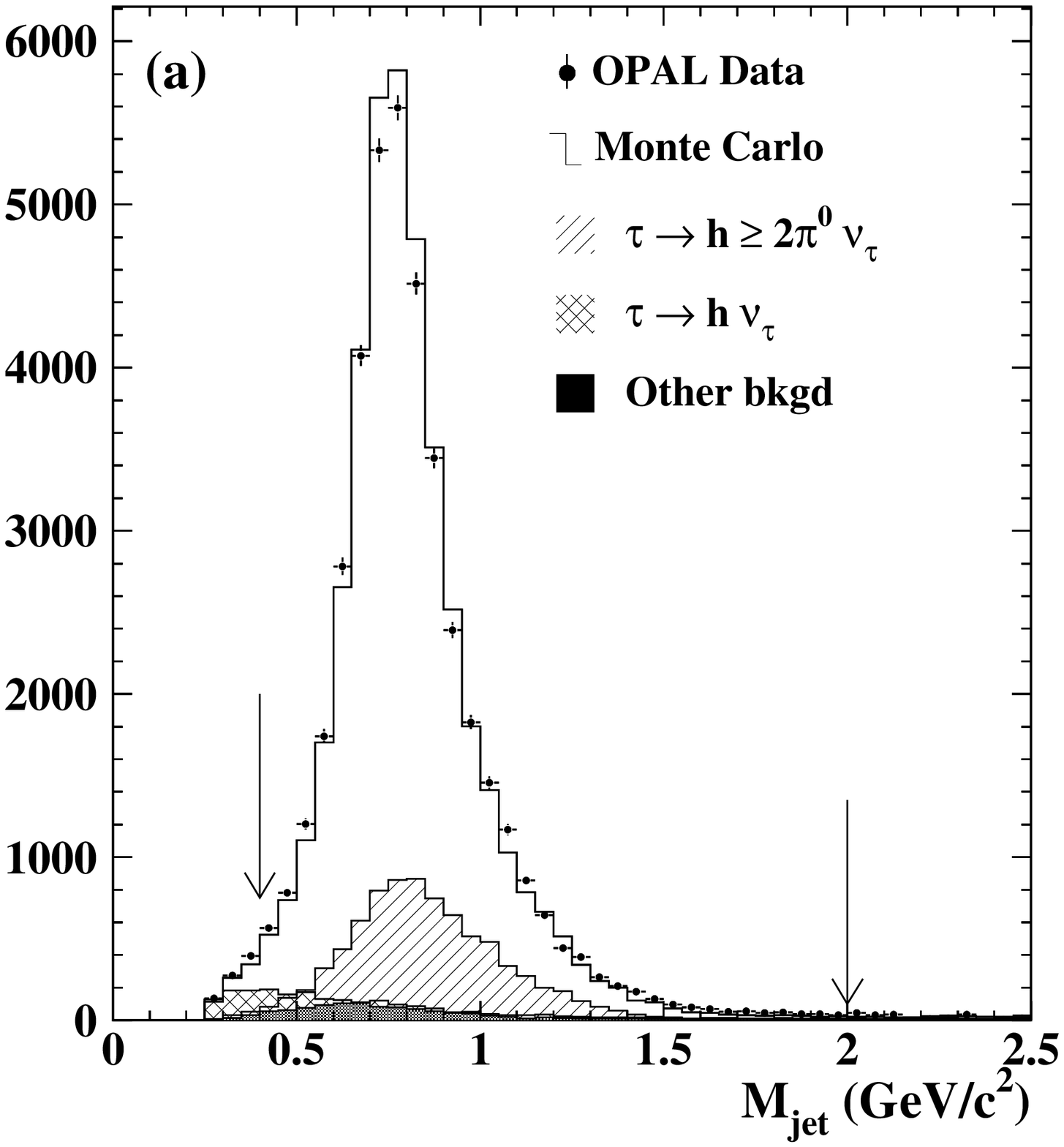,width=8.75cm,bbllx=0pt,bblly=0pt,bburx=567pt,bbury=567pt}}
\mbox{\epsfig{file=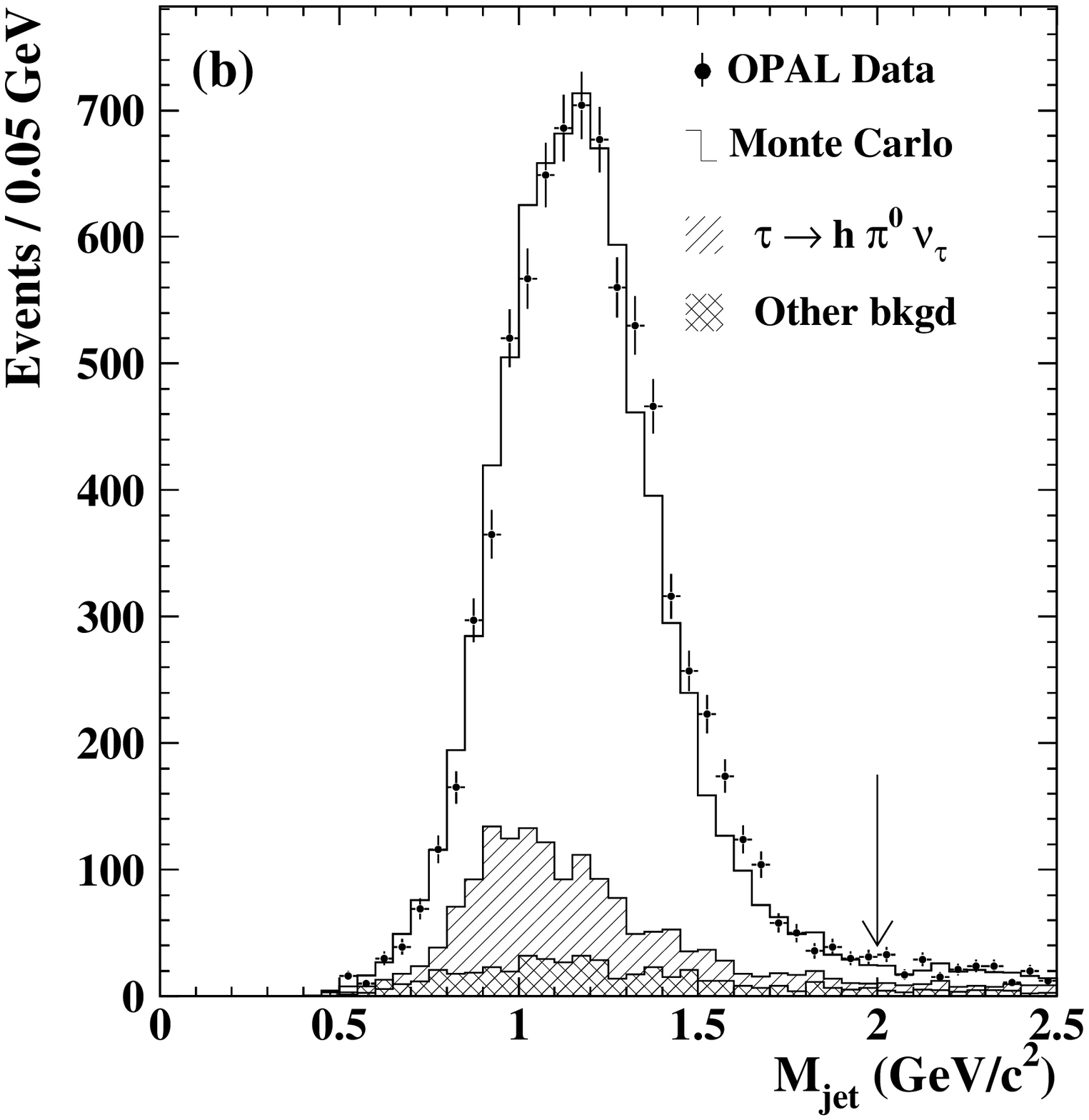,width=8.75cm,bbllx=0pt,bblly=0pt,bburx=567pt,bbury=567pt}}
\caption{\label{inv_masses_1} 
The invariant mass distribution of jets identified as \decxB\ (a)  and \decxCC\ (b). 
The mass cuts are indicated on the plots.
In figure (a) the diagonally hatched area is \decxCC\ jets, the cross hatched
area is \decxA\ jets and the dark shaded area is other background.
In figure (b) the diagonally hatched area is \decxB\ jets
and the cross hatched area is other background.}
\end{center}
\efi

% ........... Figure 7: dE/dx
\bfi
\begin{center}
\mbox{\epsfig{file=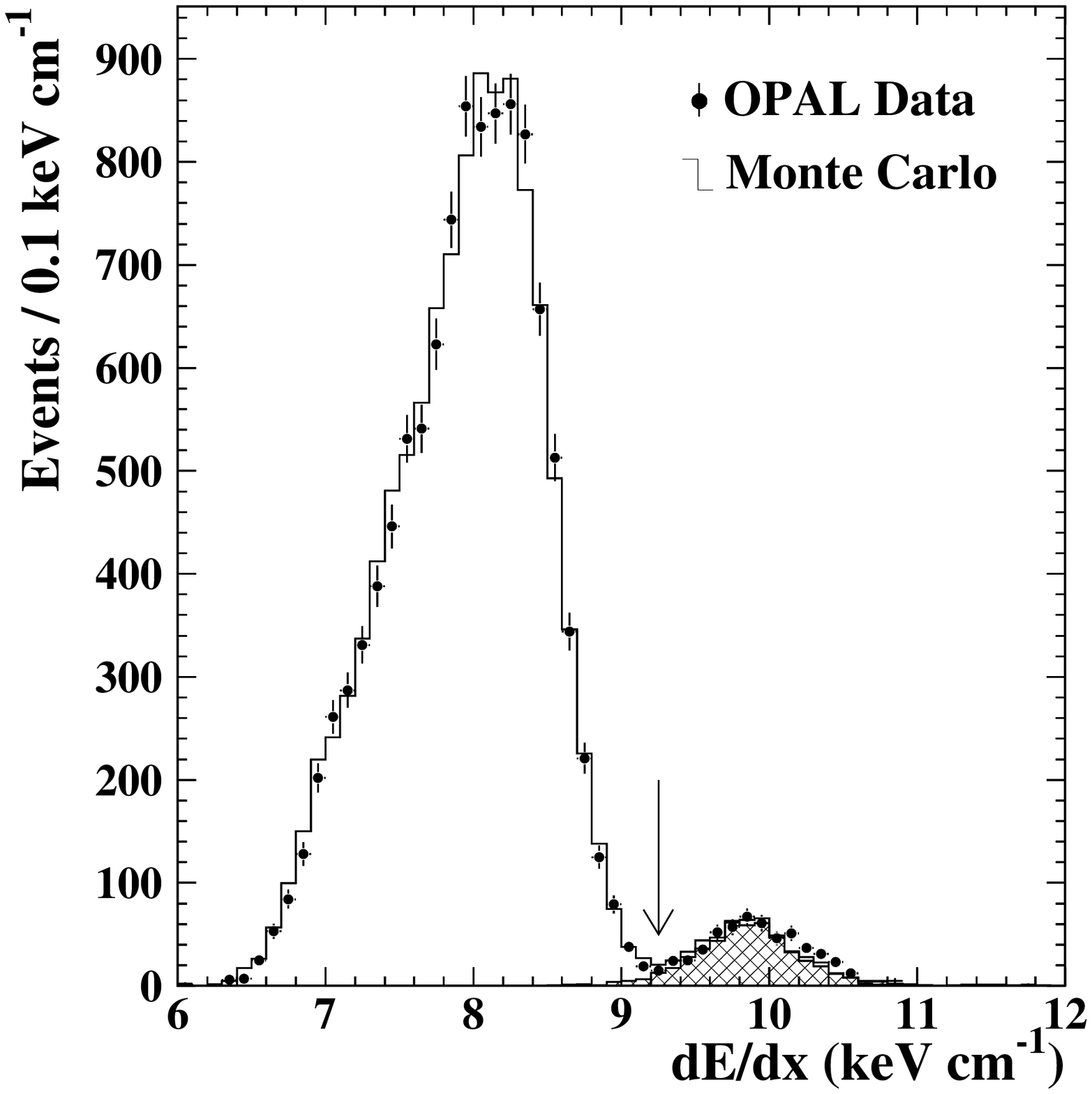,width=10.0cm,bbllx=0pt,bblly=0pt,bburx=567pt,bbury=567pt}}
\caption{\label{dedx_0pi0} The $\dedx$ distribution used to measure the \tauE\
background in the \decxB\ sample.
A background-dominated region is created by selecting jets 
with $ 9.25 < \dedx < 12.0 $~keV/cm
(indicated on the histogram). 
The cross hatched region corresponds to the \tauE\ background events.}
\end{center}
\efi

% ........... Figure: branching ratios
\bfi[t]
\begin{center}
\mbox
{\epsfig{file=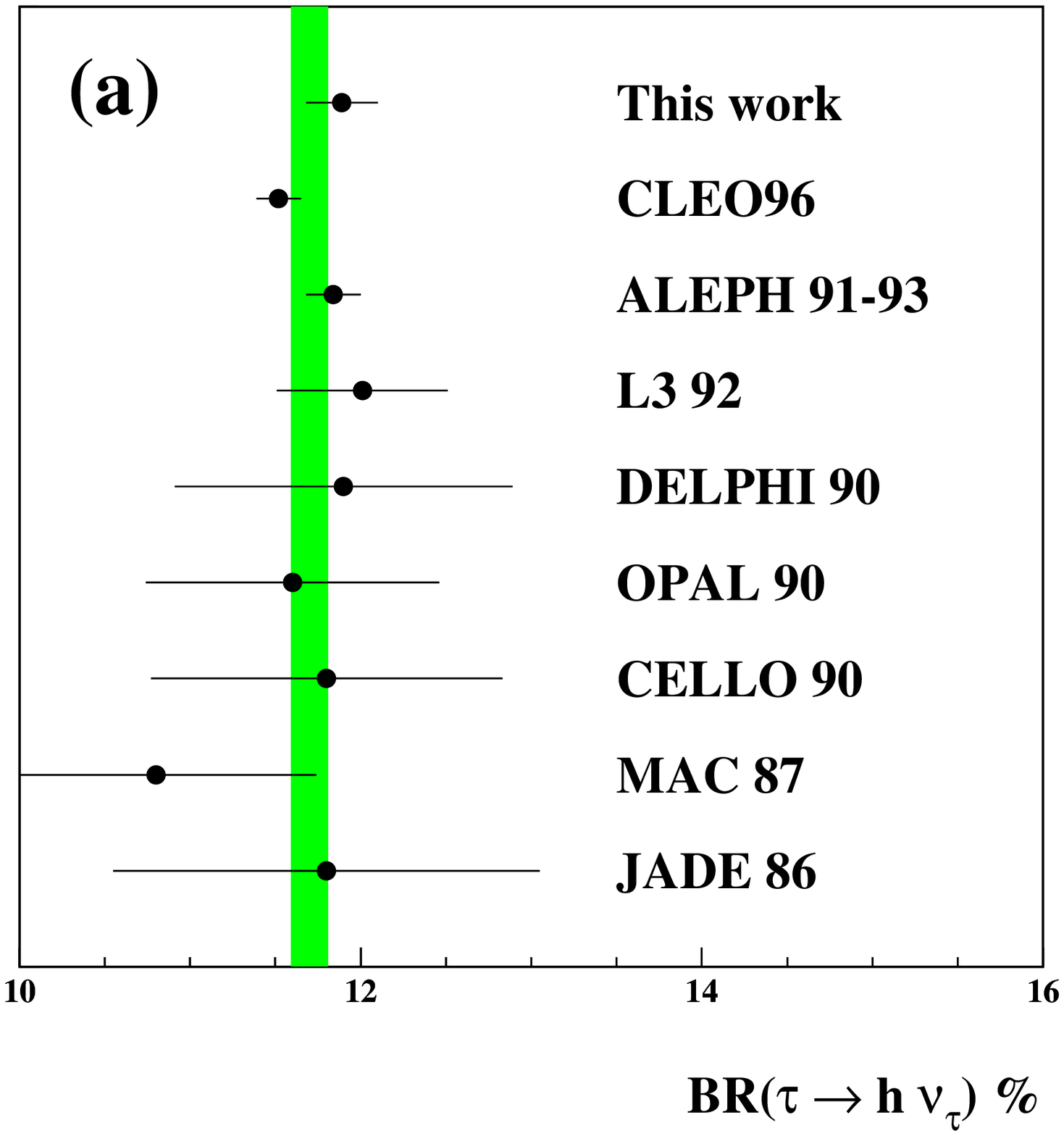,width=8.75cm,bbllx=0pt,bblly=0pt,bburx=567pt,bbury=567pt}}
\mbox
{\epsfig{file=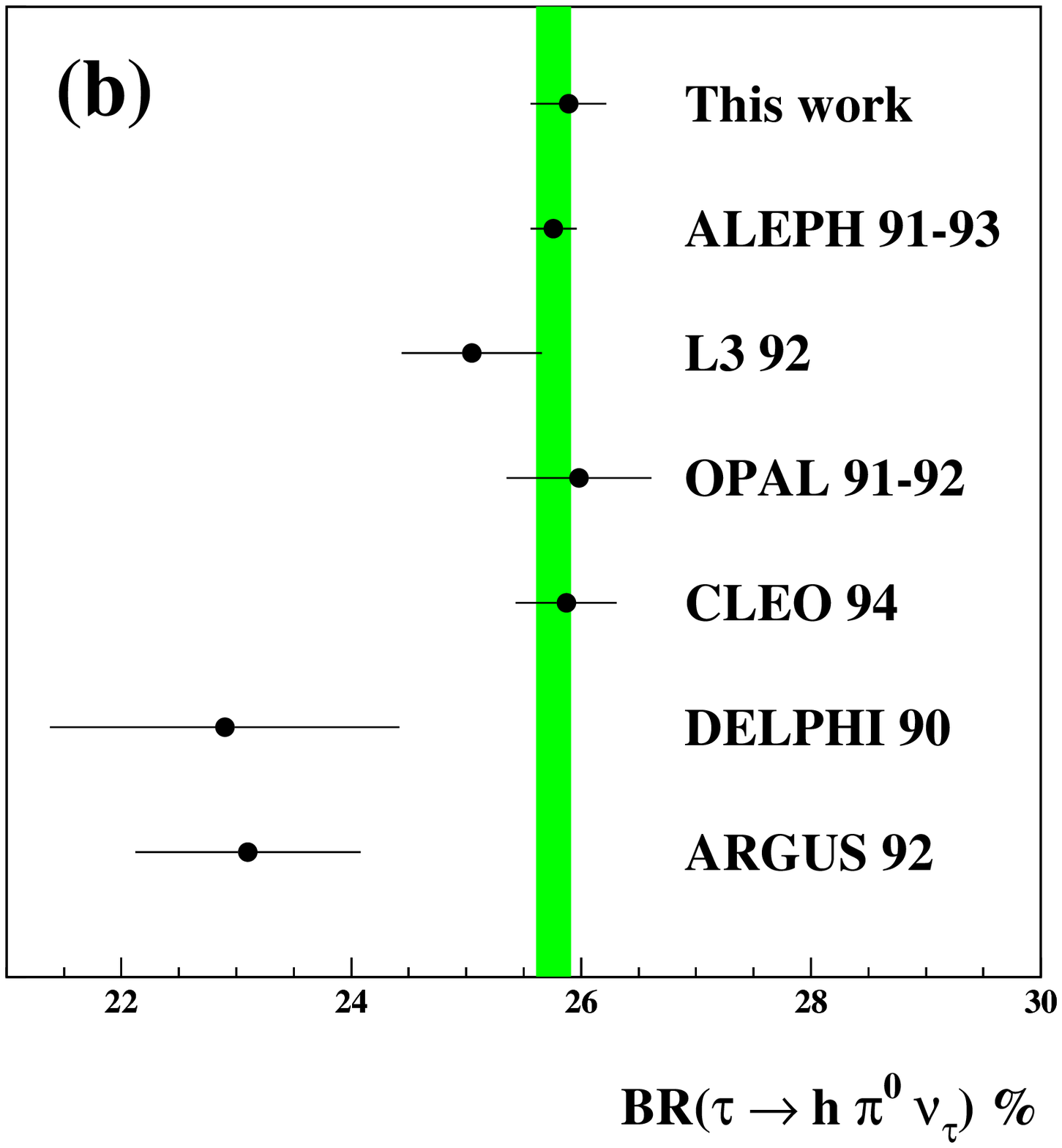,width=8.75cm,bbllx=0pt,bblly=0pt,bburx=567pt,bbury=567pt}}
\caption{\label{br_comp_0} 
The branching ratios for the \decxA\ and \decxB\ decay modes for this work
are compared with previous measurements.
The solid band in each plot is the PDG average for that channel.
The PDG average for the \decxA\ branching ratio does not include the
CLEO measurement while the
PDG average for the \decxB\ branching ratio uses all of the measurements.
The error bars shown include both systematic and statistical uncertainties.}
\end{center}
\efi

% ........... Figure: universality
\bfi[t]
\begin{center}
\mbox
{\epsfig{file=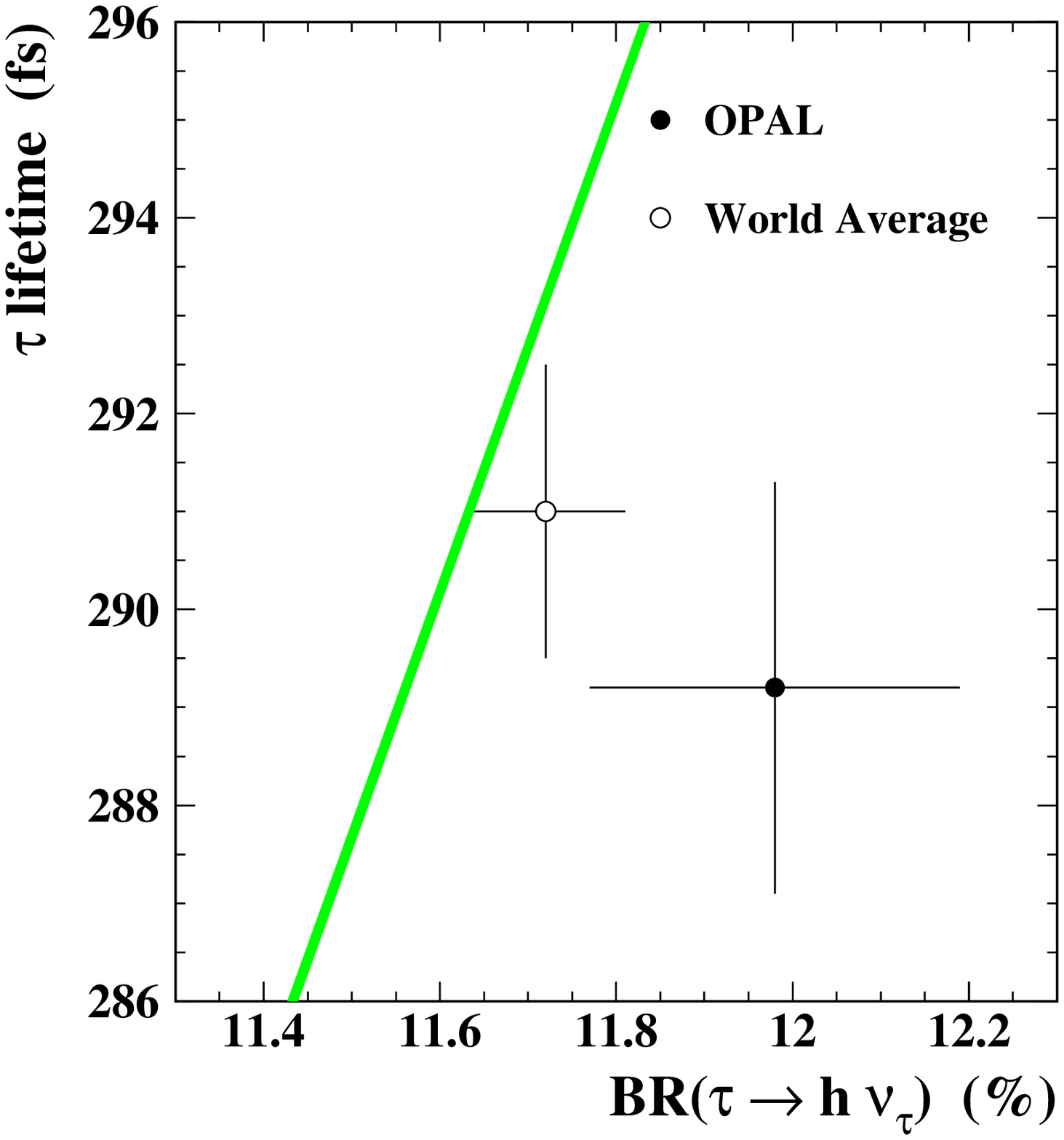,width=15.0cm,bbllx=0pt,bblly=0pt,bburx=567pt,bbury=567pt}}
\caption{\label{univ_1} 
The tau lifetime is plotted as a function of \decxA\ branching ratio.
The shaded band shows the Standard Model prediction assuming lepton
universality and its width reflects the uncertainty associated with the tau mass. 
The solid point uses the \decxA\ branching ratio obtained in
this measurement and the OPAL tau lifetime.
The open circle uses the world average for both the 
\decxA\ branching ratio and tau lifetime.}
\end{center}
\efi

\end{document}